\documentclass[fleqn,10pt]{wlscirep}
\usepackage{caption}
\usepackage[utf8]{inputenc}
\usepackage[T1]{fontenc}
\usepackage{subcaption}
\title{Observing formation and evolution of dislocation cells during plastic deformation}

\newcommand{\todo}[1]{\textcolor{blue}{#1}}

\usepackage{soul}
\usepackage{upgreek}
\usepackage{ gensymb }
\usepackage{array}
\usepackage{wrapfig}
\usepackage{afterpage}

\author[1,2]{Albert Zelenika}
\author[1]{Adam André William Cretton}
\author[3]{Felix Frankus}
\author[1]{Sina Borgi}
\author[3]{Flemming B. Grumsen}
\author[2]{Can Yildirim}
\author[2]{Carsten Detlefs}
\author[3]{Grethe Winther}
\author[1*]{Henning Friis Poulsen}
\affil[1]{Technical University of Denmark, Department of Physics, Kgs. Lyngby, Denmark}
\affil[2]{European Synchrotron Radiation Facility, Grenoble, France}
\affil[3]{Technical University of Denmark, Department of Civil and Mechanical Engineering, Kgs. Lyngby, Denmark}
\affil[*]{hfpo@dtu.dk}

\begin{abstract}
During plastic deformation of metals and alloys, dislocations self-organise in  cells, which subsequently continuously decrease in size. How and when these  processes take place has remained elusive, because observations of the structural dynamics in the bulk have not been feasible. We here present X-ray diffraction microscopy movies of the structural evolution during tensile deformation of a mm-sized aluminium (111) single crystal. The formation and subsequent development of 40,000 cells are visualised.  We reveal that cells form in a stochastic and isotropic manner already at 1\% strain. We show that the cell size and dislocation density distributions are log-normal and bi-modal Gaussian distributions, respectively, throughout. 
This insight leads to an interpretation of the formation and evolution steps in terms of universal stochastic multiplicative processes. This work will guide dislocation dynamics modelling, as it provides unique  results on cell formation.

\end{abstract}

\begin{document}

\flushbottom
\maketitle
%
%
\thispagestyle{empty}

\section{Main}

Metals and alloys are typically polycrystalline aggregates. When deformed plastically, dislocations are introduced into the lattice of each grain \cite{Taylor1934, Nabarro1967}. For materials with medium to high stacking fault energies the primary mechanism for facilitating the shape change is dislocation slip. The interplay between the plastic flow and minimisation of elastic energy implies that the dislocations organize into boundaries. Moreover the micro-structure often evolves to comprise two types of boundaries, on a coarser scale Geometrically Necessary Boundaries, GNBs, reflecting systematic variations in the plastic flow, and on a finer scale Incidental Dislocation Boundaries, IDBs, thought to represent statistically trapped dislocations\cite{Hansen2001b}.  The IDBs separate nearly dislocation-free regions called cells. 
With increasing deformation, the flow stress increases, and the entire hierarchical structure shrinks in length scale. Empirically, structural properties such as cell size and mis-orientation distributions have been shown to exhibit scaling with the applied field for plastic strains larger than 5\%-10\% \cite{Hughes1997,Godfrey2000, Hughes2000}, when separating IDBs from GNBs.    

However, despite extensive studies - motivated by the socioeconomic impact of metals - we are still unable to predict the type of micro-structure that forms as function of material and processing from first principles. While correlations have been established with the active slip systems \cite{Huang2007, Winther2007, Winther2015}, the mechanisms underlying the cell formation and cell sub-division are not known.  This prohibits realizing the vision of ``materials science in the computer'' in this field. At the root of this predicament are two issues. Firstly, the complexity and computational effort in handling the large sets of dislocations involved has been prohibitive. Discrete \cite{Bulatov2006, Kubin2013, Po2014} and Continuum \cite{Xia2015} Dislocation Dynamics models (DDD and CDD) can at best simulate representative volumes up to about 0.5 \% and 1\% strain, respectively. For that reason until recently even basic patterning has been elusive \cite{Vivekanandan2021}. 

Secondly, it is challenging to visualise the micro-structural evolution experimentally in a representative way, as the multi length scale problem requires  a combination of contrast to cells and local dislocation content over a large representative volume, \emph{in situ} and within the bulk of a sample. Traditionally, dislocation structures are mapped by electron microscopy (EM) \cite{Huang2007, Uchic2006, Echlin2012, Le2012, Choi2015, Men2023}.  EM provides very detailed maps, but is inherently limited in terms of representative volume by the use of thin foils, micro-pillars or free surfaces. Hence, the multiscale dynamics may not represent bulk conditions. For bulk studies x-ray diffraction based imaging has emerged as a powerful tool. Multi-grain X-ray imaging modalities such as 3DXRD and DCT can provide comprehensive information on the grain level, well suited for interfacing with crystal plasticity models \cite{Poulsen2003, Jakobsen2006, Pokharel2015, Juul2017, Renversade2024}. X-ray scanning nano-beam methods on the other hand have been demonstrated to visualise a few dislocations \cite{Levine2006, Hofmann2013a, Ulvestad2017}.  For quantitative measurements of dislocation densities line profile analysis of x-ray diffraction patterns prevails\cite{Wilkens1970, Krivoglaz1969, Mughrabi1986, Borbely2012, Borbély2023}, but results are averages over grains or over the illuminated volume. 

To address this multiscale experimental problem we have established Dark Field X-ray Microscopy \cite{Simons2015,Simons2018}, DFXM. With modalities similar to dark field TEM \cite{Williams2012}, this method enables large field-of-view visualisations of both dislocations \cite{Jakobsen2019, Simons2019, Dresselhaus2021, Yildirim2023} and cells\cite{Simons2015, Zelenika2024} within the bulk.  We here apply the technique to a comprehensive study of the structural evolution within Aluminum during \emph{in situ} tensile deformation from an applied strain of $\epsilon = 0$ to  $\epsilon = 0.046$. The inspected volume comprises $\sim$ 40,000 dislocation cells, providing excellent statistics in this initial patterning range, where the structure is little known and the overlap with the strain regime available in simulations is most prominent.

\begin{figure}[hbt!]
    \centering
    \includegraphics[width=0.95\linewidth]{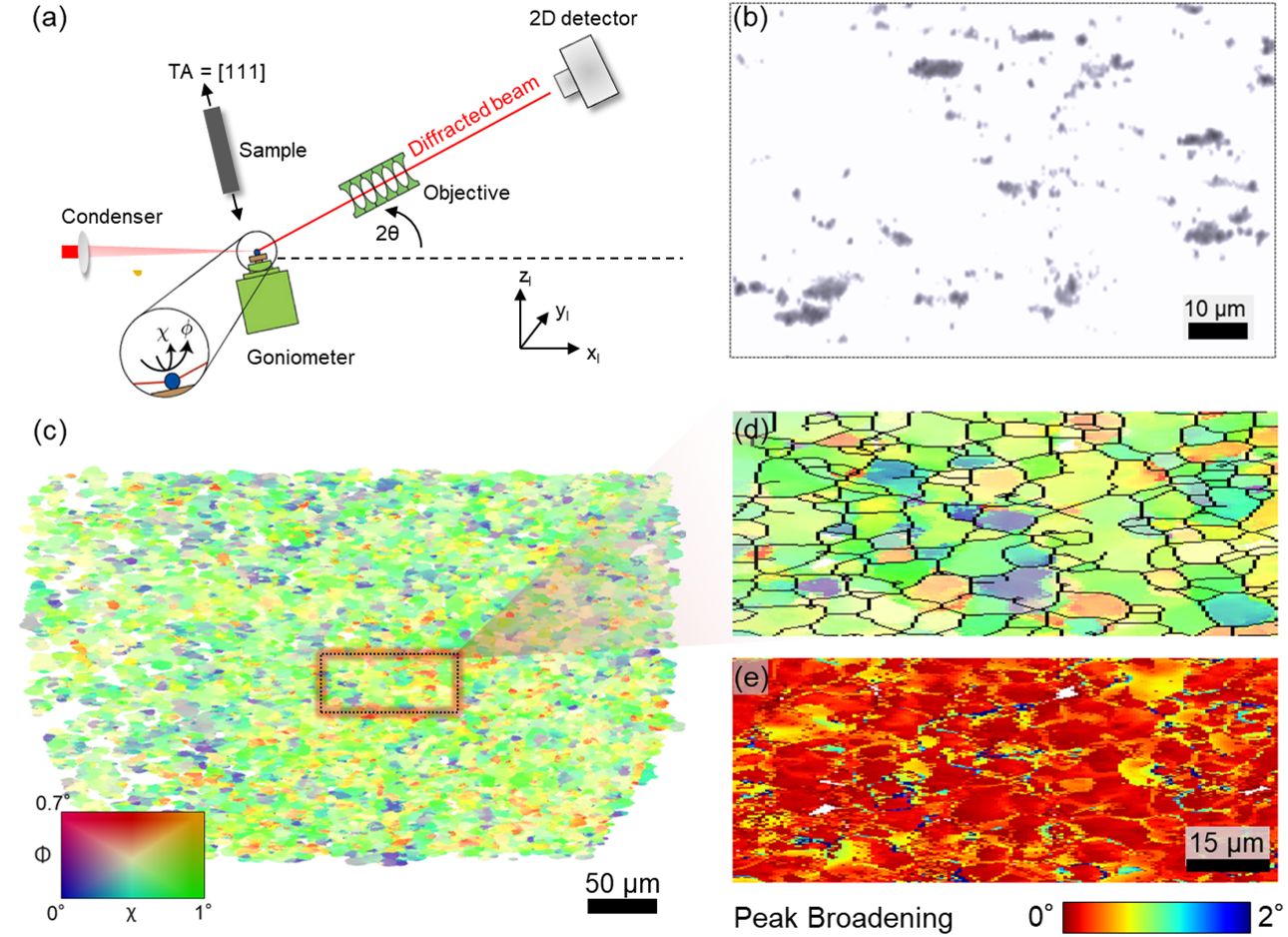}
    \caption{ \textbf{DFXM Geometry and examples of maps}. a) Sketch of DFXM set-up. The incoming beam is condensed in $z_l$  thereby illuminating one ($x_{l}, y_{l}$)-layer in the specimen. The diffracted signal is magnified by the objective. The tensile axis (TA) of the sample is aligned with the diffraction vector. b) Raw image in weak beam mode revealing dislocations and dislocation entanglements for $\epsilon=0.002$. c) Orientation map for $\epsilon = 0.046$ with an Inverse Pole Figure color code. d) Zoom-in on region marked in c) with misorientations above 4$^{o}$ as black lines. e) Corresponding map of the average peak broadening in directions $\phi$ and $\chi$, indicative of the total dislocation density. } 
    \label{fig:overview}
\end{figure}

The sample of choice is a single crystal  of pure Aluminum, with [111] parallel to the tensile axis, see Fig.~\ref{fig:overview} a).  The mechanical device is mounted in the DFXM microscope such that diffraction in the vicinity of the [111] reflection is imaged for the illuminated layer. See Fig.~\ref{fig:stress-rig} for more details of the loading device. By tilting goniometer angles $\phi$ and $\chi$, see Fig.~\ref{fig:overview} a), contrast is provided to orientation and strain components. Three complementary modalities are illustrated in Fig.~\ref{fig:overview} b)-e). Weak beam contrast provides statistics over dilute dislocations ensembles for $\epsilon \le 0.005$, orientation contrast provides statistics over cell properties, once these have formed, while the peak broadening $\Delta q$ is a proxy for the total dislocation density, $\rho$: $\Delta q \sim \sqrt{\rho}$.  By repeating the mapping for a set of $z_l$ layers, a 3D map can be created. For definitions, algorithms and specifications, see sections \ref{sec_DFXM_methodology} and \ref{sub_theory_peakbroadening}.

 Single crystals of the present orientation form parallel planar GNBs in addition to cells \cite{Kawasaki1980}. The GNBs are clearly manifest in both EM and DFXM when inspecting planes which include the TA (see Fig.  \ref{fig:OIM_other_plane} and Zelenika \emph{et al.} \cite{Zelenika2024}). In this work we report on the cell evolution primarily within a plane perpendicular to the TA. Here the GNBs are known to be less visible \cite{Kawasaki1980}. This was confirmed by the absence of preferred directions or orientation correlations between cells in the present data, see Figs.~\ref{fig:misorientation_analysis} and \ref{fig:anisotropy_ROI}. This indicates that a model that does not explicitly take the crystallography of the deformation process (slip planes) into account may be adequate.

For $\epsilon < 0.01$, the sample exhibits a set of isolated dislocations and dislocation entanglements, as evidenced by Fig. ~\ref{fig:overview} b). In Section \ref{sub-pristine} these clusters are quantified in three complementary ways, which sample the elastic field associated with the dislocations differently. Consistently the three approaches reveal that the clusters are randomly positioned and distributed homogeneously within the Field-Of-View (FOV). Moreover their size distributions are consistent with log-normal distributions.

\begin{figure}[hbt!]
    \centering
    \includegraphics[width=0.85\linewidth]{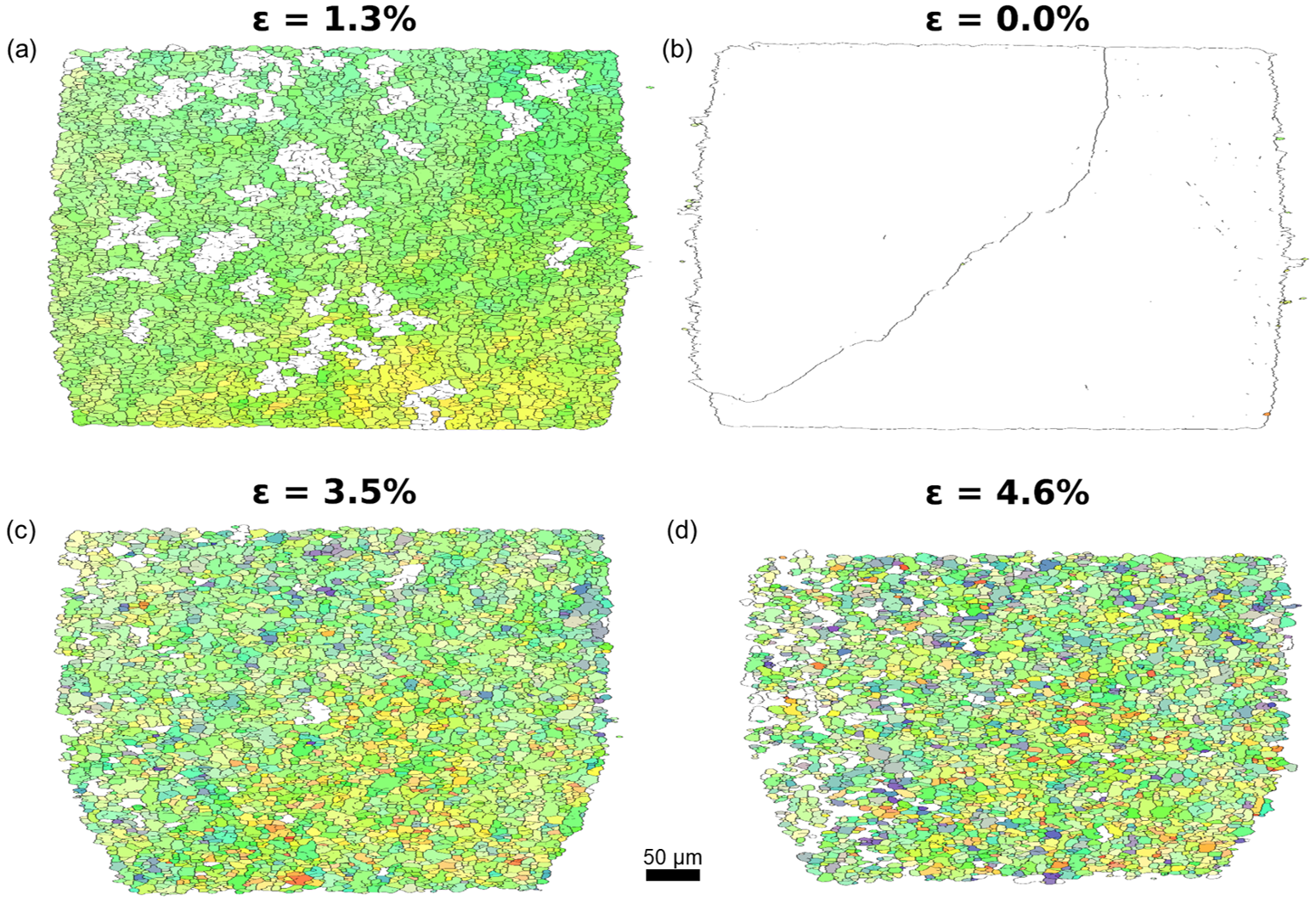}
    \caption{\textbf{Cell formation.} a)-d) Cells defined by a misorientation filter for selected applied strains, $\epsilon$, within the Field-of-View of the microscope and for one of the 9 layers inspected. The color is indicative of the average orientation of each cell. }
    \label{fig:cell_formation}
\end{figure}
\vspace{-2mm}

\begin{wrapfigure}{l}{0.41\textwidth}
  \centering
  \includegraphics[width=0.89\linewidth]{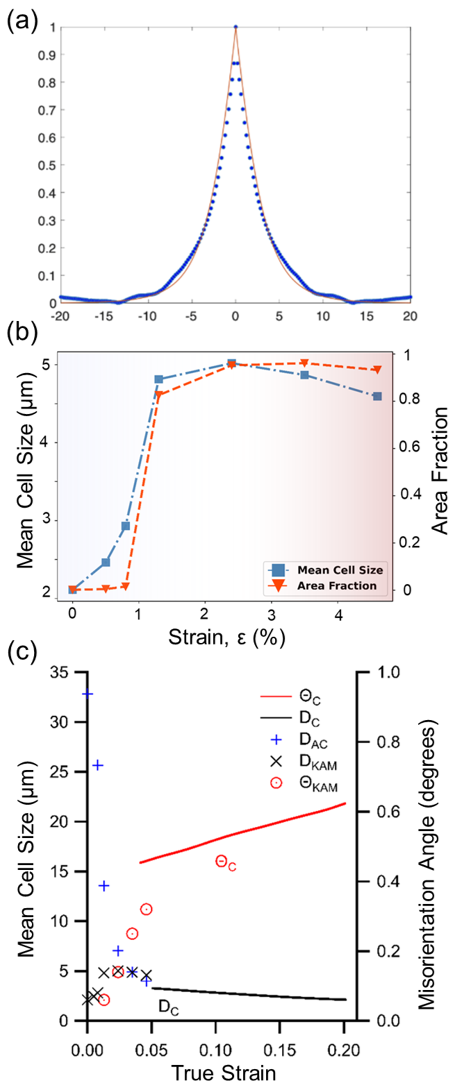}
  \vspace{-1mm}
  \caption{
  \textbf{Structural order.} a)  Autocorrelation of the orientation map for $\epsilon= 0.046 $ (blue points) along $y_l$ with a superposed model based on assuming the cells to be hard spheres (red line). All parameters in the model are provided by the experimentally determined cell size distribution. b) Area fraction of cells and average cell size as function of $\epsilon$. c) Comparison of the properties measured by DFXM (symbols) with literature values based on TEM \cite{Hansen2001} (lines). For the average cell size  the literature values  ($D_C$)  are compared to the results of the cell analysis ($D_{KAM}$) and of the autocorrelation analysis
 ($D_{AC}$). For average misorientation the literature values ($\Theta_C$) are compared to the cell analysis values ($\Theta_{KAM}$).}
  \label{fig:cell_properties}
\end{wrapfigure}

With increasing applied strain, the orientation spread and the total dislocation content - both averaged over the entire volume inspected - grow approximately linear with the strain, see Fig.~\ref{fig:macroscopic_widths}. 
Moreover, domains of approximately uniform intrinsic orientation distribution appear, see Supplementary Video 1. We define cells as domains where the Kernel Average Misorientation, KAM, of all boundary voxels are above a threshold, $\theta_{\textrm{KAM}}$.  By means of morphological operations, the boundaries are made 1 pixel thin, and as a consequence a tessellation of the sample is obtained, see  Fig.~\ref{fig:KAM_mask_details}. Anticipating scaling we set $\theta_{\textrm{KAM}}$ to be a linear function of $\epsilon$. The resulting KAM filter superposed on the orientation map are shown as function of $\epsilon$ as Supplementary Video 4. Snapshots are provided in Fig.~\ref{fig:cell_formation}. Inspection reveals that the tessellations are indeed similar, justifying the linear ansatz.  This analysis also allows us to address a long-standing question: \emph{when are the cells formed?} As shown in Fig.~\ref{fig:cell_properties} b) the sample undergoes a  transformation from no cells to site-saturation of cells in the range $\epsilon = 0.008 $ to $\epsilon = 0.023 $. From Fig.~\ref{fig:cell_properties} c) it appear that this screening of the surrounding takes place as soon as the cells are fully formed, at $\epsilon = 0.024$. Moreover, it is shown that results are consistent with TEM results for $\epsilon > 0.05$. 

In a complementary approach we study the order in the structure on the micrometer length scale by deriving the autocorrelation function of the orientation maps. These are  presented as function of $\epsilon$ in Supplementary Video 3. The absence of any side peaks clarifies that there is no long range ordering  of cells.  Next in Fig.~\ref{fig:cell_properties} a)  for $\epsilon = 0.046$,  we compare the autocorrelation function with a hard sphere model of the cells with all parameters defined by the experimentally determined size distribution, see also Suppl. Information. The excellent match with experimental data informs that the cells ``do not see each other''. We interpret this as evidence of the local and random nature of the cell formation process. 

\begin{figure}[hbt!]
    \centering 
    \includegraphics[width=0.9\linewidth]{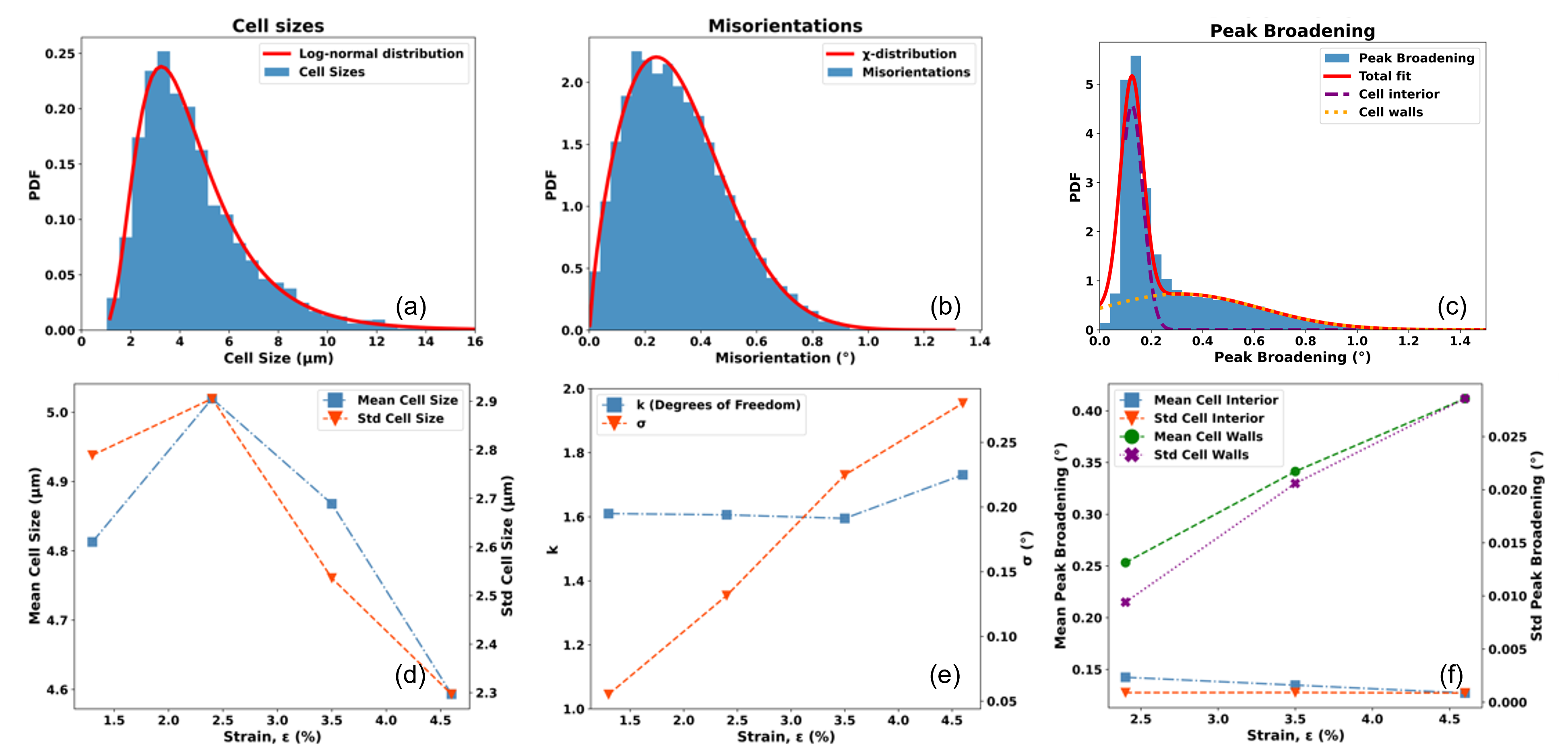}
   \caption{ \textbf{Statistical properties of the set of cells and their evolution with applied strain.} Above: distributions for the 40,000 cells at $\epsilon=0.046$. a) Cell size distribution with best fit to a log-normal distribution overlaid (red line). b) Misorientation angle distribution with best fit to a $\chi$-distribution superposed (red line). c) Distribution of the average peak broadening in $\phi$ and $\chi, \Delta q$, for each voxel. This is a proxy for the square root of the total dislocation density. A best fit to two Gaussians is superposed (red line). These are referred to as cell interior and cell walls. Below d)-f): corresponding  evolution in the model parameters with with applied strain. In e) k and $\sigma$ are the two parameters in the chi distribution.  } 
    \label{fig:evolution}
\end{figure}

The cell map at  $\epsilon = 0.046$ comprises $\approx 40,000$ cells within the 9 layers analysed - one of them shown in Fig.~\ref{fig:cell_formation} d). This large ensemble allows us to distinguish with certainty between functional forms for various distributions, thereby strongly constraining models of structural evolution. The cell size distribution is found to be consistent with a log-normal distribution, see Fig~\ref{fig:evolution} a), while statistical tests reject other commonly used functions, cf. Fig.~\ref{fig:grainsize_testmodels}. Likewise, the distribution of misorientation angle between neighboring cells  is well described as a $\chi$ function, Fig~\ref{fig:evolution}  b).  The corresponding peak broadening distribution, shown as Fig~\ref{fig:evolution}  c),  is a proxy for the local \emph{total} strain, assuming that the dislocation density approximately reflects the plastic strain. The distribution is bimodal, which we interpret as an elastic strain and a plastic strain component. The former is to a good approximation normal, within data uncertainty the latter may be normal or log-normal. As expected the plastic strain component is predominantly present in the rather broad cell walls, cf. Fig.~\ref
{fig:overview} e). Distributions of aspect ratio are provided in Supplementary Information.  
\vspace{-3mm}

Next, with the statistical tools presented, we describe the  micro-structural evolution.  Within the applied strain range from $\epsilon = 0.013$ to 0.046, where the material goes from predominantly formation (increasing cell sizes) to predominantly fragmentation (decreasing cell sizes), the distributions evolve in a consistent way, as summarised in  Fig.~\ref{fig:evolution} d) to f). Specifically, within statistical error the size distribution is log-normal throughout and exhibit scaling ($\sigma$ is constant.) Likewise, the misorientation angle distribution grows linear with strain and exhibits scaling (within experimental error k is constant).  Moreover, the bimodal model for $ \Delta q$ is valid throughout. The area fractions of cell interior  and wall regions remain approximately constant, as do the elastic distribution for the cell interior. In contrast the wall distribution widens in a linear fashion with $\epsilon - \epsilon_0$, with $\epsilon_0 = 0.012$ being the onset of cell formation, cf. Fig.~\ref{fig:evolution} f).


The existence of a log normal cell size distribution throughout is consistent with \emph{both} cell formation and division being multiplicative stochastic processes. This finding leads to suggest that the combined cell formation and division process can be described as a Markovian growth-fragmentation process\cite{Bertoin2017}. Used e.g. in population science \cite{Gyllenberg2007} and chemical engineering \cite{Bosetti2020}, such processes are mathematically proven to give rise to log-normal distributions and scaling. Specifically, our understanding is as follows:  individual dislocation entanglements appear randomly in time and space. Similar to particle creation by diffusion they grow with a growth rate that is proportional to their size.
 Following impingement, the cell pattern exhibits scaling with $\theta_{\mathrm{KAM}} \sim \epsilon$. In the cross-over from formation to subdivision their area remains about constant, while new dislocations continue to build up the boundaries, the IDBs. The linear growth in dislocation content leads to a linear growth in the average misorientation across the IDBs. With the cells exhibiting no long range order, the IDB statistics can be modelled as a sum of stochastic processes, consistent with  misorientations being associated with a chi distribution\cite{Pantleon1996}.  Finally, during fragmentation the larger cells are more likely to divide. This is corroborated by a positive correlation between cell heterogeneity and cell size, cf. Fig.~\ref{fig:corr_strain_and_gradient_inside_cells}.

In the past, log-normal-distributed plastic strain \cite{Tang2020}, misorientation \cite{Gurao2014} and geometrically necessary dislocation densities \cite{Jiang2013} have been obtained experimentally with surface studies at the \emph{grain} scale for polycrystalline metals deformed to strains exceeding $\epsilon=0.1$ and  across a range of materials and operative micro-mechanical mechanisms like slip and twinning. Log-normal distributions have also been predicted for the plastic strain component by crystal plasticity simulations \cite{Chen2021} of \emph{grain }ensembles, without any dislocations and for cell size and misorientation angle by assuming a cell splitting probability proportional to the cell surface area and misorientation evolution by rotational diffusion \cite{Sethna2003}.  

Uniquely x-ray imaging has the angular resolution that provides contrast for cells at all relevant applied strain levels. The DFXM maps shown here represent a representative volume deeply embedded in a sample sufficiently large that the mechanical conditions represents bulk conditions. 
The maps presented can be input to crystal plasticity FEM or - in future - CDD models. Subsequently two 3D movies may be compared: one experimental and one with the simulations. As demonstrated in the past e.g. for grain growth such a comparison of hundreds of thousands of points in space-time will provide unprecedented opportunities for guiding and validating models \cite{Zhang2020}.    The methodology presented applies to poly-crystals and all crystallographic space groups, limited mainly by the spatial resolution of 100 nm.  To study the individual formation and fragmentation events we are commissioning a new goniometer and a more mechanically stable loading device, making it possible to track the micro-structural features in space and time by DFXM. Moreover, this may be complemented by mapping the local (purely elastic) longitudinal strain component \cite{Ahl2017}. Such extensive mapping is also relevant for studies of other processes such as annealing (recrystallisation) and ductile damage, as it will combine mapping of the deformed microstructure with large volume identification of nuclei and voids, respectively.

\section{Methods}

\subsection{Sample}  The sample is a single crystal of 99.9999 \% pure Aluminum of dimensions $1 \times 1 \times 20$ mm. The tensile axis is (111). After cutting, the sample was annealed at 540 $^\circ$C for 10 hours. 

The sample was mounted by glue on a grooved PEEK holder with a gauge length of 5 mm. This  was inserted in a four-point bending loading device with the sample on the tensile side. In this geometry the sample is subject to uniaxial tension. The tensile device is illustrated in Fig.~\ref{fig:stress-rig} and the resulting stress-strain curve in Fig.~\ref{fig:stress-strain}. 

\subsection{DFXM experiment}

The DFXM experiments were conducted at Beamline ID06-HXM at the European Synchrotron Radiation Facility, ESRF. For details of the set-up see Kutsal \emph{et al. } \cite{Kutsal2019}. A monochromatic beam with an energy of 17 keV was focused to a line with a FWHM of $\approx 600 \,\mathrm{nm}$ in the $z_{\ell}$ direction, illuminating a layer within the sample. The scattering angle for the Al \{111\} Bragg reflection is  $2\theta = 17.98 ^\circ$. The objective was a Be Compound Refractive Lens (CRL), with 88 lenslets with radius of curvature $ R = 50 \mu$m, positioned at a sample-to-CRL-entry distance of $d_1 = 269$ mm and a CRL-exit-to-detector distance of $d_2 = 4987$ mm.
The corresponding magnification and numerical aperture are $\mathcal{M} = 18.52$ (measured) and $NA = 0.705$ mrad (calculated from Eq.~9 in Poulsen \emph{et al.} (2017) \cite{Poulsen2017}),  respectively. The 2D detector was located 5256 mm from the sample. With an additional magnification of 2 in the detector the effective pixel size was  656 nm (along $x_l$) $\times$ 202 nm (along $y_l$). The corresponding field of view in the sample is $350\,\mathrm{\mu m} \times 900\,\mathrm{\mu m}$. The scan parameters used are listed in Table 1 in Supplementary Information. 
The exposure time for a single image was 0.2 second including motor movements.

We did not observe any creep in the microstructure over the time of the acquisitions (minutes to hours). 

\subsection{Data analysis}
With the exception of the weak beam data, the entire analysis is based on the output of 
darfix \cite{Ferrer2023}, as shown in Supplementary Videos 1 and 2. Specifically, for each voxel a 2D Gaussian is fitted to the $(\phi,\chi)$ distribution. The subsequent analysis is detailed in Supplementary Information. 

\section{Data availability}
All the data presented in this paper, along with the analysis tools used for data evaluation, are available on GitHub at https://github.com/adcret/PMP. These resources include all data sets and scripts used to process and analyze the data following initial treatment steps with the darfix software. A comprehensive description of the analysis can be found in the "$cell\_refinement\_analysis.ipynb$", further details about the packages and dependencies of this analysis tool are found in the "$README.md$" file. Data DOI: \hyperlink{doi.org/10.15151/ESRF-ES-776857198}{doi.org/10.15151/ESRF-ES-776857198} 

\bibliography{references}



\section*{Acknowledgements}

We thank Kristian M\o lhave for suggesting the design for the tensile rig. This work was funded by the European Research Council (Advanced grant no 885022) and by Danish Agency for Science and Higher Education (grant number 8144-00002B). C.Y. acknowledges the financial support by the ERC Starting Grant "D-REX" (no 101116911).We acknowledge ESRF for a PhD grant and for the provision of synchrotron radiation facilities under proposal number MA-4442 on beamline ID06-HXM. 

\section*{Author contributions statement}

HFP wrote the manuscript with significant contributions from AZ and AC, then edits and correspondence from all co-authors. The conceptual idea of the study is due to GW and HFP.
The planning and execution of the experiment was lead by AZ, with contributions from all co-authors except AC. Primarily AZ but also AC performed data analysis under the supervision of HFP, GW and CY. FF, FG and CD designed and built the four point bender. 

\section*{Additional information}
The authors of this work have no competing interests in this work.

\section*{Supplementary information} 

\textbf{Supplementary Information}
Supplementary Notes I and Figs. S1–\ref{fig:compare_size}.
\vspace{0.5cm}

\noindent \textbf{Supplementary Video 1: Center-of-Mass $(\phi,\chi)$ orientation maps}. Left: the full field-of-view for the central layer with a region-of-interest, ROI, marked by the rectangular box. Right: a zoom-in corresponding to the ROI. The color scale employed varies from strain level to strain level and is indicated by inverse pole figures inserted. 
\vspace{0.5cm}

\noindent \textbf{Supplementary Video 2: Peak broadening maps}. The voxel-by-voxel width (FWHM) of the avearge peak broadening $\Delta q = \sqrt{\Delta q_\phi^2 + \Delta q_\chi^2}$, with $\Delta q_\phi$ and $\Delta q_\chi$ being the FWHM's resulting from a 2D Gaussian fit to the $(\phi,\chi)$-distribution. Left: the full field-of-view for the central layer with a region-of-interest, ROI, marked by the rectangular box. Right: a zoom-in corresponding to the ROI. 
\vspace{0.5cm}

\noindent \textbf{Supplementary Video 3: Auto-correlation maps}. Center lines of the 2D autocorrelation function for the central layer along directions $x_l$ and $y_l$. 
\vspace{0.5cm}


\noindent \textbf{Supplementary Video 4: Center-of-Mass $(\phi,\chi)$ orientation maps with KAM mask overlaid}. The images are replica of those in Supplementary Video 1 with a Kernal-Averaged-Misorientation mask (black lines) overlaid. 
\vspace{0.5cm}
    
\noindent \textbf{Supplementary Video 5: Call maps.} Cells (colored regions) are identified as connected regions using a Kernel-Averaged-Misorientation (KAM) mask (black lines). The colors of the cells represent their average orientation, as indicated by the inverse pole figures inserted. Left: the full field-of-view for the central layer with a region-of-interest, ROI, marked by the rectangular box. Right: a zoom-in corresponding to the ROI. 
\vspace{0.5cm}
    
\noindent \textbf{Supplementary Video 6: Peak broadening maps with KAM masks overlaid}. The images are replica of those in Supplementary Video 2 with the KAM mask (black lines) overlaid.
\vspace{0.5cm} 

\newpage
\section*{Supplementary information}
\renewcommand{\thefigure}{S\arabic{figure}}
\setcounter{figure}{0}
\subsection{DFXM methodology}
\label{sec_DFXM_methodology}

The geometry of DFXM, coordinate systems, associated diffraction formalism and interfacing to micro-mechanical modelling is presented in detail in Poulsen \emph{et al.}, 2017 \cite{Poulsen2017} and Poulsen \emph{et al.}, 2021\cite{Poulsen2021}, while the details of the instrument at ID06-HXM, ESRF are presented in Kutsal \emph{et al.}, 2019\cite{Kutsal2019}. For convenience we summarise the definitions and optical properties relevant for this article.

\begin{figure}[hbt!]
    \begin{center}
    \resizebox{0.8\columnwidth}{!}{\includegraphics{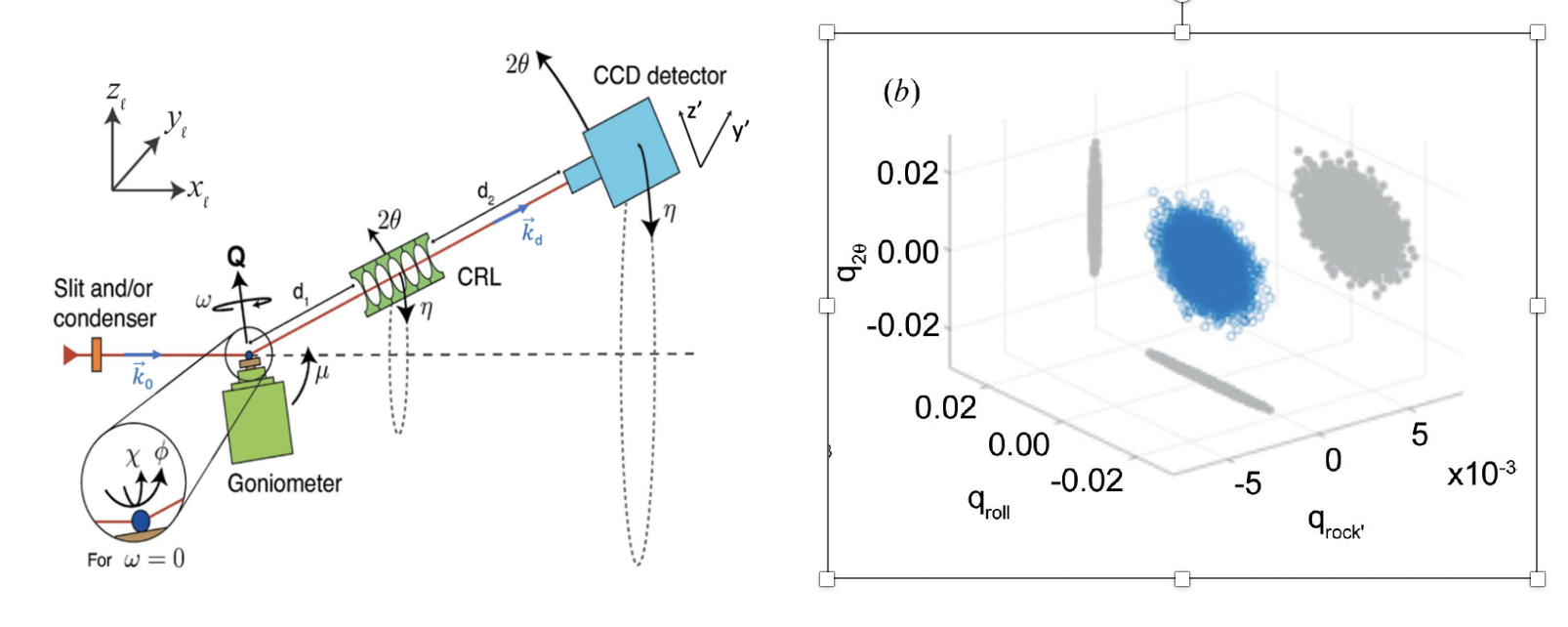}}
    \end{center}
\caption{ a) DFXM geometry as shown in the laboratory coordinate system, $(x_\ell, y_\ell, z_\ell)$. The optical axis before the sample corresponds to the average incident beam direction. The optical axis of the objective (a Compound Refractive Lens\cite{Snigirev2009}) is co-linear with the average diffracted ray, when the sample is in the nominal position for the Bragg peak. The pivot point of the goniometer and sample is coincident with the intersection of the two optical axes. Vector  $\vec{Q}$ defines the local scattering vector at a given point $\vec{r}(x,y,z)$ within the sample. The scattering angle is $2\theta$. The goniometer has four tilts, $\phi$,$\chi$, $\omega$ and $\mu$. Only the former two is used in this work. b) Reciprocal space resolution function with units of strain, represented as a 3D scatter plot (blue).  For ease of interpretation, the resolution function is projected onto the planes $\vec{q}_{rock} =0, \vec{q}_{roll} =0,$ and $\vec{q}_{2\theta} =0$ (all grey) \cite{Poulsen2021}. 
}
\label{fig:DFXM-principle}
\end{figure}

The setup is illustrated in Fig.~\ref{fig:DFXM-principle} a). A nearly monochromatic beam  illuminates the sample. This beam is condensed in the vertical direction to generate essentially a line beam. The goniometer is designed to access diffraction angles in a vertical scattering geometry, and probe reciprocal space in the immediate vicinity of a given reflection, here  $\vec{Q} = (1,1,1)$.
The optical axis of the objective is aligned to the diffracted beam to produce an image on the 2D detector.  The objective acts as a classical microscope magnifying an objective plane within the sample onto the detector plane. This microscope is characterised by the numerical aperture, $NA$, and the focal distance, $f_N$ of the objective, as well as the magnification  of the x-ray signal, $\mathcal{M}$, and the Field-of-View, FOV, in the sample plane. Due to the layered incoming beam, the DFXM images are affine transformations of the structure in the $(x_{\ell},y_{\ell}$)-plane with an effective pixel size that is $1/\tan(2\theta)$ larger along $x_{\ell}$ than $y_{\ell}$. To generate 3D maps the sample is translated relative to the incident beam in direction $z_\ell$ and acquisition is repeated.

In the simplified geometry used in this work, the goniometer has two angular degrees of freedom, the two orthogonal tilt directions $\phi$ and $\chi$, with $\phi$ corresponding to  rotation around $y_{\ell}$. Diffraction contrast is obtained by scanning the sample in these two directions, known as rocking and rolling scans, respectively. These scans probe two shear components out of of the 9 components of the micro-mechanical tensor field \cite{Poulsen2021}. As always, the two shear components comprise both a rotational and an elastic strain part. We cannot separate the two, but the magnitude of the field values implies that the rotation part is more dominant for the larger applied strains.
 
 To probe the field of the axial strain, the "$2\theta$-arm" - comprising both the objective and the detector - may be rotated corresponding to a shift of $\Delta 2\theta$. However, this modality was not applied in this work. Instead the scans performed are mosaicity scans, 2D grid scans in $\phi$ and $\chi$, repeated for a number of layers. As a result each voxel in the illuminated part of the sample are associated with a 2D distribution of intensity values, a function of $\phi$ and $\chi$.  

 A 2D Gaussian model to this distribution provides in general good fits to this intensity distribution. As a result five parameters are derived for each voxel:
 \begin{itemize}

 \item \emph{the Center-of-Mass, COM, in $\phi$ and $\chi$}. We present these values in terms of poles in the (111) polefigure. Moreover, we quantify the evolution in the mosaicity map with $\epsilon$ in terms of texture evolution, by  setting the third orientation axis, not measurable from only one diffraction vector, to a constant. 

\item \emph{the normalised peak widths, $\Delta q_{\phi}$ and $\Delta q_{\chi}$}, in units of strain. These are the widths (FWHM) of the fitted Gaussian in the two directions. To reduce complexity these results are combined by calculating the scalar width 

\begin{eqnarray}
\Delta q = \sqrt{ \Delta q_{\phi}^2 + \Delta q_{\chi}^2 }. \label{eq-peakwidth_scalar}
\end{eqnarray}

\item \emph{the amplitude}. In kinematic diffraction theory this is proportional to the fraction of the voxel that is illuminated. In practice this is also influenced by vignetting and by uneven sampling in reciprocal space caused by the objective.

\end{itemize}

Notably, the reciprocal space part of the experimental resolution function of the microscope is very asymmetric. A Monte Carlo simulation is provided in Fig.~\ref{fig:DFXM-principle} b) \cite{Poulsen2021}, representative of the setting used here. The width (FWHM) of the resolution function in direction $\chi$ is defined by the NA of the objective, here 0.705 mrad.  In contrast the width of the resolution function in direction $\phi$ is dominated by the incoming divergence of the beam. This is 10 times smaller, of order 0.1 mrad. In comparison the local angular spread within the prestine sample was determined to be 2.0 mrad and 0.77 mrad, respectively. Moreover, the data acquisition algorithm involved a continuous scanning over $\phi$ where intensities  are integrated over 0.04 deg = 0.69 mrad. Hence, it appears the finite angular resolution can be neglected for most of the work below.

\subsection{Relationship between local peak broadening and dislocation density}
\label{sub_theory_peakbroadening}

The relation between the broadening of diffraction peaks and dislocation density is well studied when it comes to \emph{the longitudinal direction}. Within the limitation of a dilute system the longitudinal strain can be described in terms of the dislocation density, $\rho$, as\cite{Borbély2023}   
\begin{eqnarray} 
 \Delta q_{long} & = \frac{\Delta d}{d} =\Big[ \frac{b^2}{4\pi} C_g \ln \big( \frac{R_e}{r_0}\big) \Big]^{1/2} \sqrt{\rho}. \label{eq-elastic}
\end{eqnarray}
In this equation there are three constants: $b$, the modulus of the Burgers vector and $R_e$ and $r_0$, the outer and inner cut-off of the dislocation system, respectively. The contrast factor $C_g$  depends on the specifics of the active dislocation systems, but will be approximately constant during loading for the process studied here. This peak width purely reflects the elastic response. 

In contrast broadening along the two \emph{shear directions} will be combinations of elastic and plastic contributions, for higher applied strains dominated by the plastic spin part (rotation of the lattice). 
In Electron Back-Scatter Diffraction, EBSD, the misorientation between 
neighboring pixels is used to derive the geometrically necessary dislocation content \cite{Pantleon2008}. The analysis can be seen as a generalisation of the simple relationship between the misorientation angle $\theta_{mis}$ across a dislocation wall comprising identical and equi-spaced edge dislocations with neighboring distance $a$ and corresponding density $\rho$. For small  $\theta_{mis}$
\begin{eqnarray} 
\theta_{mis}/2  = b/a \sim b\sqrt{\rho}.  \label{eq-plastic}
\end{eqnarray}
The same procedure applies to DFXM, where it is possible to derive all tensorial components, provided a 3D map is acquired.  However, the spatial resolution function has to be taken into account. 

Thanks to its high angular resolution, DFXM offers another modality, based on the \emph{local peak broadening in the shear directions} $\phi$ and $\chi$. For each voxel a local ($\phi$,$\chi$) distribution is acquired. This informs of the total dislocation configuration. A full exposure is outside the scope of this article. Here we will rely only on the two second moments of the distribution, the normalised peak widths $\Delta q_{\phi}$ and $\Delta q_{\chi}$ introduced above. From these we define the scalar average normalised peak $\Delta q$ by Eq.~\ref{eq-peakwidth_scalar}.
By analogy to Eqs.~\ref{eq-elastic} and \ref{eq-plastic} we make the ansatz that the measured $\Delta q$ is proportional to $\sqrt{\rho}$  and that the proportionality constant is independent of $\epsilon$ within the range explored here. 

It is at times of interest to determine a proxy for the dislocation density within a local region, e.g. in the vicinity of a cell boundary.
The Gaussian approximation applied implies that

\begin{eqnarray} 
\Delta q^{\mathrm{region}}  = \Big( \sum_{i \; \in \; \mathrm{region}} (\Delta q_i)^2  +   KAM_{i}^2 \Big)^{1/2} .  \label{eq-peakwidth_region}
\end{eqnarray}

To provide absolute numbers for dislocation densities, it is required to use data acquisition schemes involving more than one scattering vector. However, we argue that the arguments and conclusions presented are not restricted by this limitation. In particular we note that for stochastic process, the central-limit-theorem - and its extension to multiplicative processes - implies that contributions from different dislocation families (probed by different scattering vectors) add in a way that conserves the shape of the distributions.  

\subsection{Load frame and stress-strain curve }

\begin{figure}[hbt!]
    \centering
    \includegraphics[width=0.7\linewidth]{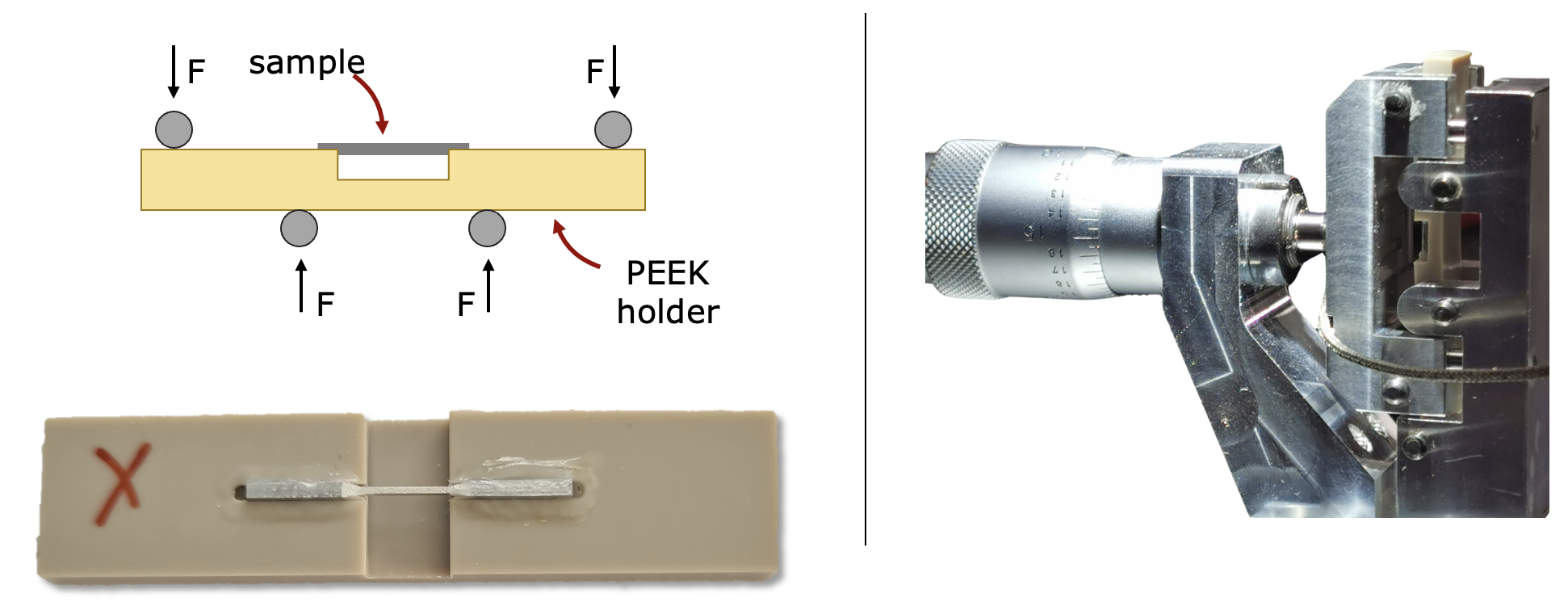}
    \caption{Principle of the tensile loading device. Left: sketch of four point bender illustrating the mount of the sample holder in a PEEK holder. Right: Realisation of the loading device.}
    \label{fig:stress-rig}
\end{figure}

\begin{figure}[hbt!]
    \centering
    \includegraphics[width=0.7\linewidth]{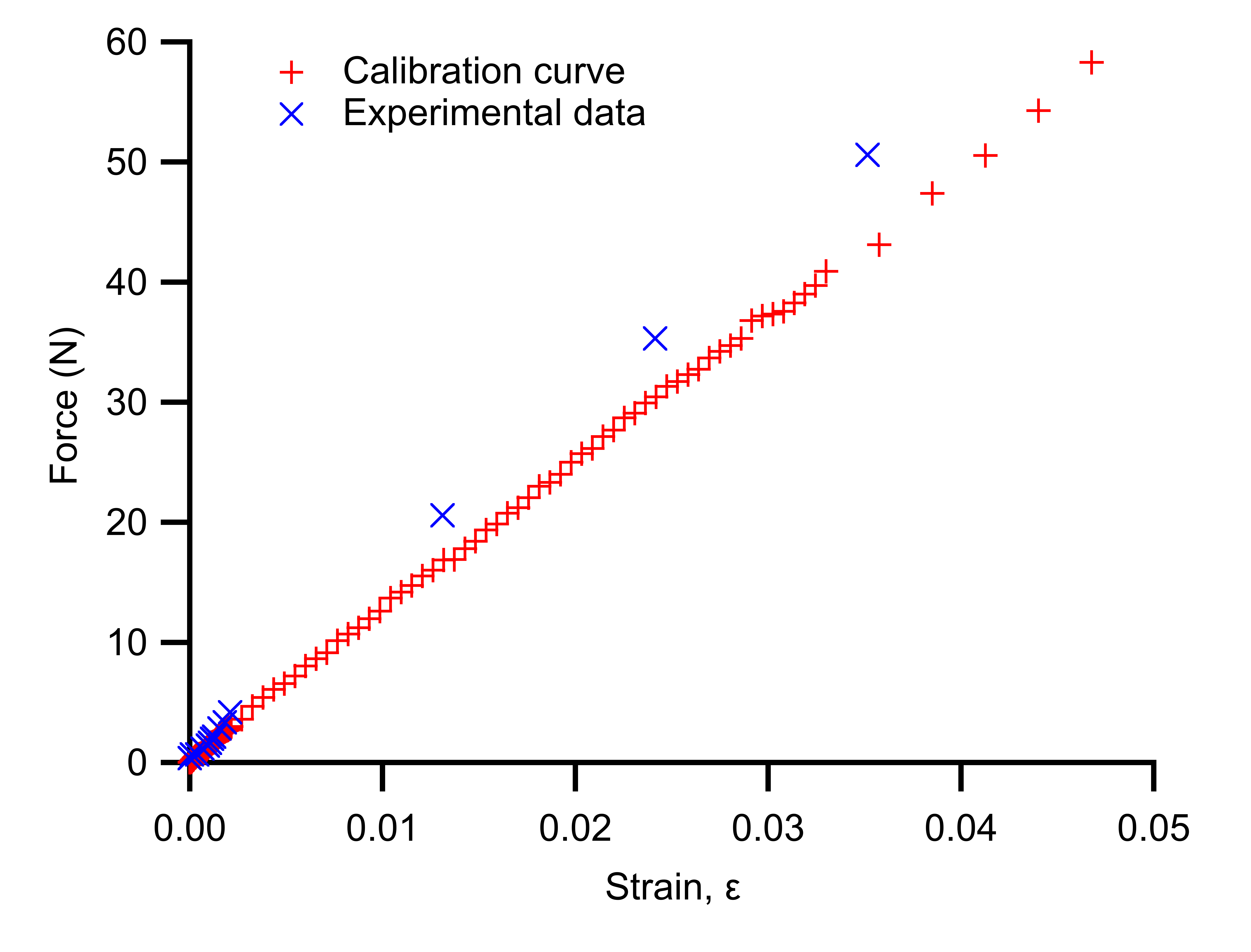}
    \caption{Stress-strain curves. Experimental data (blue) are compared with a calibration curve on a similar sample.}
    \label{fig:stress-strain}
\end{figure}

The specimen is glued to the upper surface of a notched polymer holder that is subject to bending, see Fig.~\ref{fig:stress-rig}. Due to the small cross-section of the sample compared to the polymer holder and its position far away from the neutral axis, the load case is considered to be uniaxial tension. The bending is imposed through linear deflection of the two upper rolls of the setup using a micrometer screw. A "calibration" of the axial macroscopic strain in the sample and deflection in rig through the micrometer screw has been performed using an optical light microscope and DIC, see Fig.~\ref{fig:stress-strain}.

Between each DFXM scan the applied strain was increased manually. The strain rate is of order $10^{-4}$/s. In Fig.~\ref{fig:stress-strain} we display the resulting relation between the force acted on the four point bender and the strain. Apparently, the force is nearly linear in the strain above the yield point.

\subsection{Data acquisition scheme and overview of data analysis}

\begin{table}[hbt!] 
\centering 
\renewcommand{\arraystretch}{1.3}
\begin{tabular}{|c| >{\centering\arraybackslash}p{3.0cm} |>{\centering\arraybackslash}p{3.0cm}|>{\centering\arraybackslash}p{3.5cm}|}
    \hline 
    \textbf{Applied strain} & \textbf{Nr mosalayers and optical magn.} & \textbf{Spacing in z} & \textbf{Scan ranges}  \\
    \hline 
    0 & 1 in 2x and 16 in 10x & 1 $\mu$m & 0.6\degree in 30 steps in $\phi$ and 0.5\degree in 10 steps in $\chi$\\
    0.002 & 16 in 2x and 16 in 10x & 3 $\mu$m for 2x and 1 $\mu$m for 10x & 0.56\degree in 14 steps in $\phi$ and 0.56\degree in 14 steps in $\chi$  \\
    0.005 & 16  in 10x & all 3 $\mu$m & 0.56\degree in 14 steps in $\phi$ and 0.56\degree in 14 steps in $\chi$ \\
    0.008 & 25 in 10x & 9 with 5 $\mu$m , 5 with 3 $\mu$m, 11 with 10 $\mu$m & 0.6\degree in 15 steps in $\phi$ and 0.6\degree in 15 steps in $\chi$ \\
    0.013 & 11 in 10x & all 10 $\mu$m & 0.68\degree in 17 steps in $\phi$ and 0.8\degree in 20 steps in $\chi$ \\
    0.024 & 11 in 10x & all 10 $\mu$m & 0.88\degree in 22 steps in $\phi$ and 0.96\degree in 24 steps in $\chi$ \\
    0.035 & 11 in 10x & all 10 $\mu$m & 0.88\degree in 22 steps in $\phi$ and 0.96\degree in 24 steps in $\chi$ \\
    0.046 & 18 in 10x & 3 with 5 $\mu$m, 11 with 10, 4 layers with 10 $\mu$m, interlaced with the 11 $\mu$m & 0.88\degree in 22 steps in $\phi$ and 0.96\degree in 24 steps in $\chi$ \\
    \hline 
\end{tabular}
\caption{List of mosaicity scans performed. }
\label{tab:scans}
\end{table}

Apart from weak beam data acquired for $\epsilon \le 0.002$   the scans performed during the \emph{in situ} experiment are all mosaicity scans: two-dimensional scans of $\phi$ and $\chi$. The scan parameters are listed in Table~\ref{tab:scans}. 

Due to small mis-alignments between load steps it is not the exact same set of layers that is mapped. Hence, tracking is not feasible with the mechanical device used, but statistical comparisons and tests are. To improve the statistics the analysis performed for $\epsilon =  0.002, 0.005$ and 0.046 are based on 9 layers. (Tests proved the results for the individual layers to be identical within statistical error.) For reasons of computer resources, the analysis for the  other strain step are based on the middle layer only. The experimental data for all layers and all strain steps are available in the metadata.

Most of the data analysis is based on the initial use of darfix for each layer. darfix generates pole-figures, cf. section~\ref{sub-texture} and $(\phi,\chi)$-distributions for each voxel. As described in section \ref{sec_DFXM_methodology} the latter are generally well described as 2D Gaussian distributions. The resulting center-of-mass orientation maps and peak broadening maps are provided as Supplementary Videos 1 and 2, respectively. Note that for $\epsilon = 0.035$ and $0.046$ there are minor voids in the maps, caused by a lack of intensity, as these parts had orientations slightly outside the $(\phi,\chi)$-range mapped. Next cells are identified by the use of a Kernel-Averaged-Misorientation filter, see section \ref{sec:cell_variable} and Supplementary Video 5.

This work is supplemented by an analysis of ordering on $\mu$m length scale, presented in section \ref{sec-autocorrelation}.

\subsection{Macroscopic evolution }
\label{sub-texture}

\begin{figure}[hbt!]
    \centering
    \includegraphics[width=0.6\linewidth]{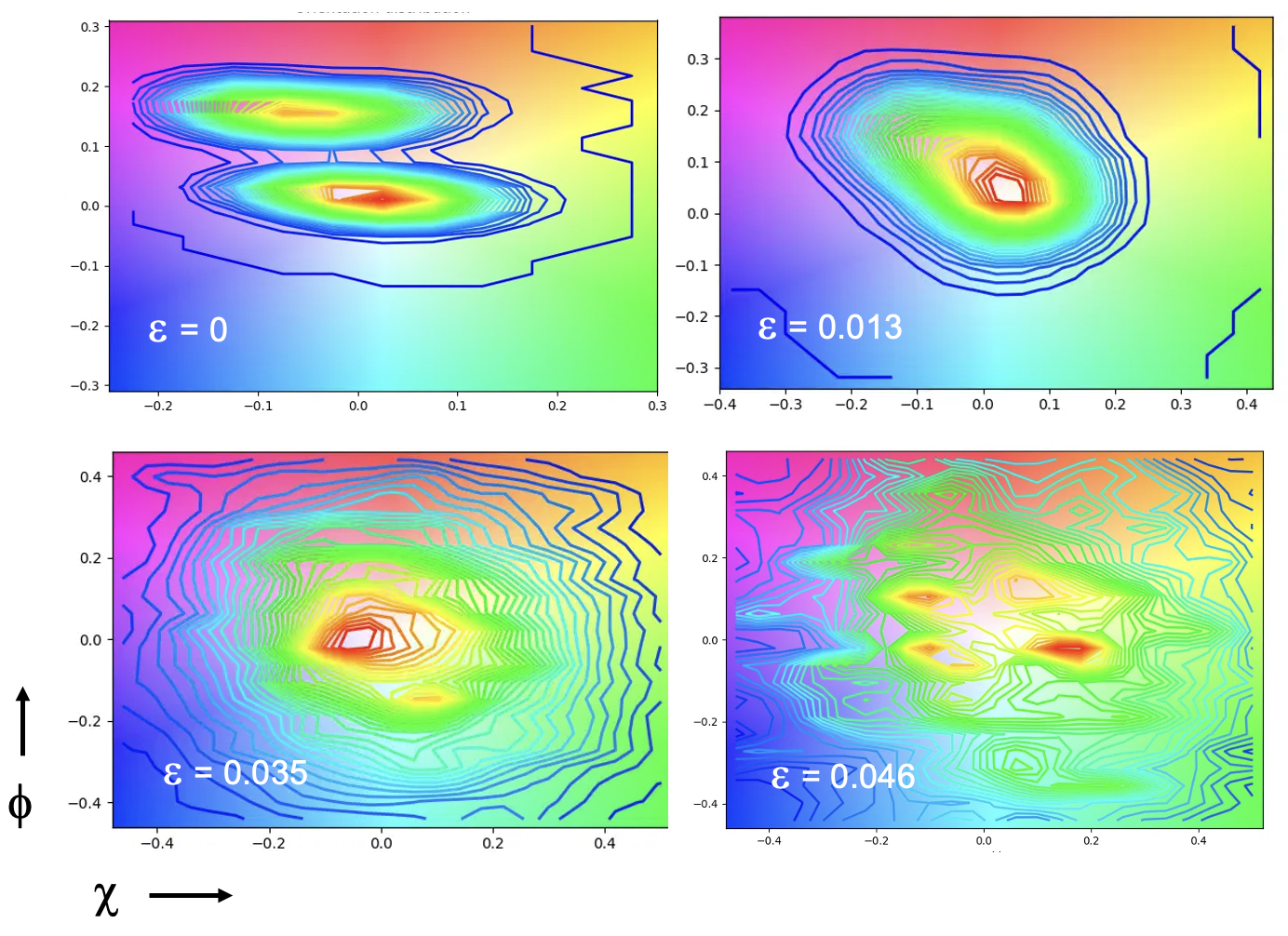}
    \caption{ Pole figures for 4 applied strain levels with varying angular ranges.}
    \label{fig:polefigures}
\end{figure}

\begin{figure}[hbt!]
    \centering
    \includegraphics[width=0.4\linewidth]{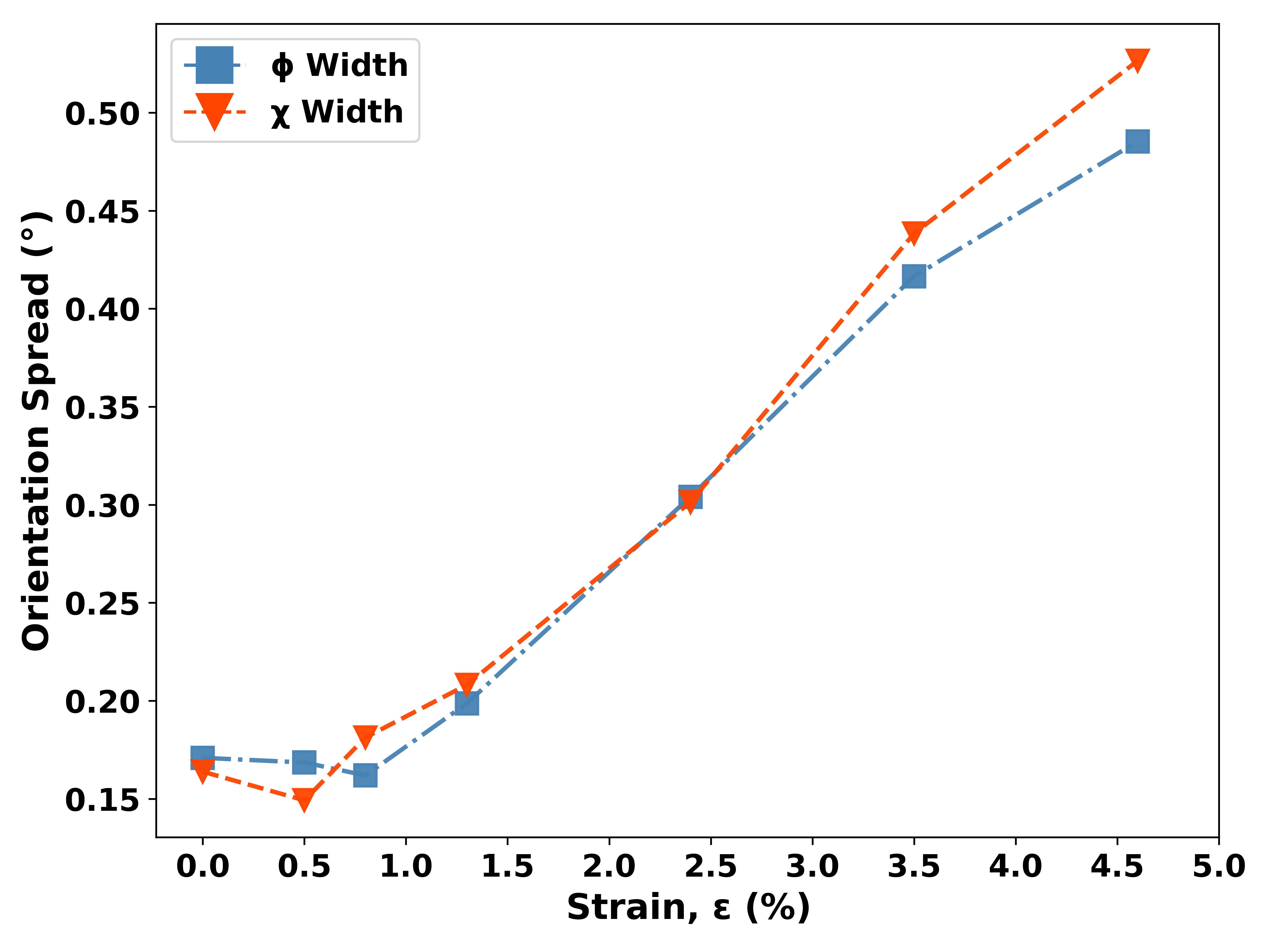}
    \includegraphics[width=0.4\linewidth]{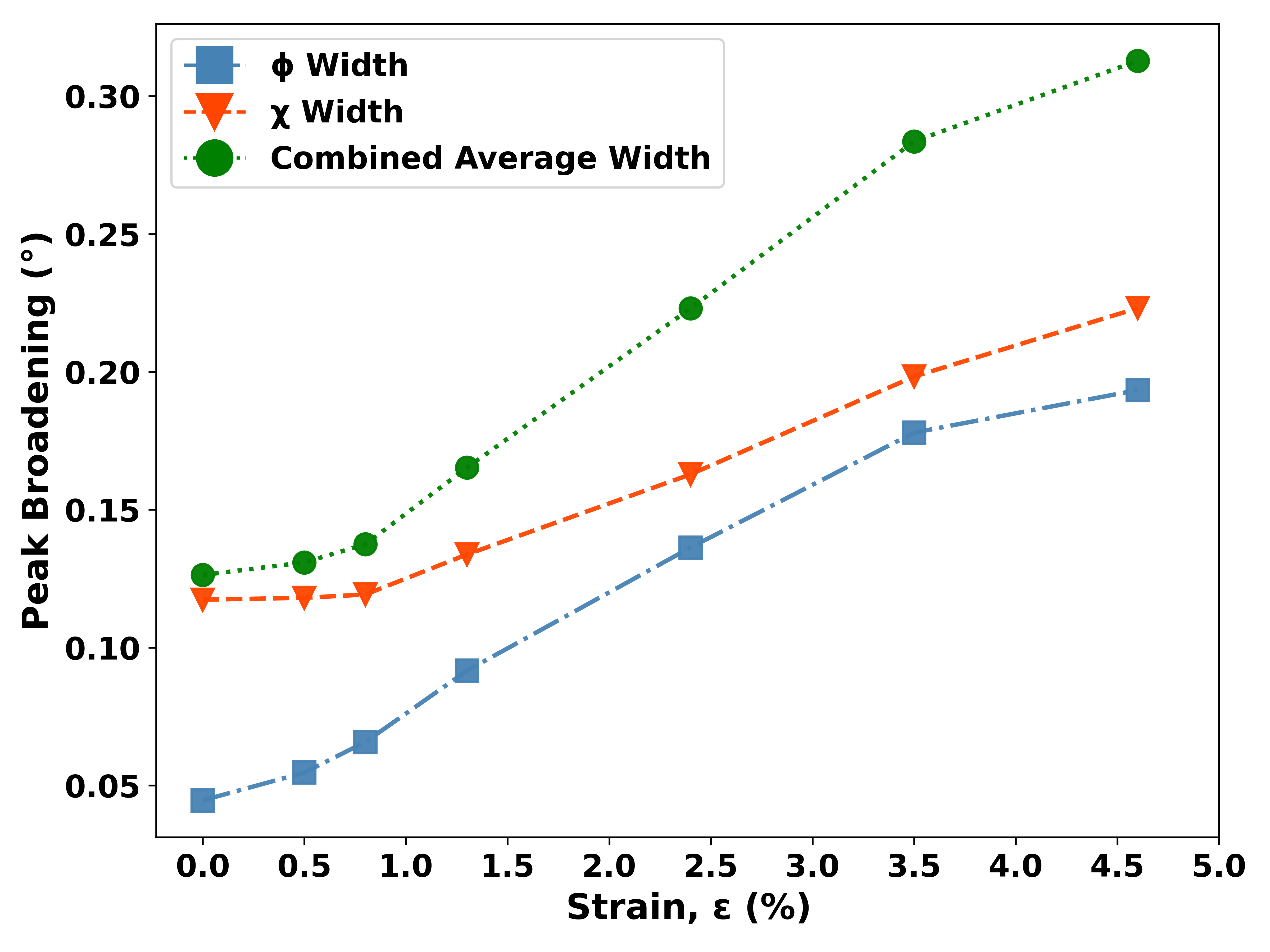}
    \caption{ Macroscopic properties as function of applied strain.  a) The width (FWHM) of the polefigure. b) Average values for the local peak broadening (FWHM). }
    \label{fig:macroscopic_widths}
\end{figure}
Initially we report on the evolution of structural properties when averaged over all voxels. Shown in Fig.~\ref{fig:polefigures} are the resulting \emph{Pole figures} for the middle layer for strain steps $\epsilon = 0, 0.013, 0.035$ and 0.046. This output was generated by darfix\cite{Ferrer2023}. It represents the summed intensity over all images. In the undeformed state the sample is a bi-crystal, with a mis-orientation between two domains of $\sim 0.15 \deg$. At higher applied strain levels the distribution becomes an approximately isotropic Gaussian.

Shown in Fig.~\ref{fig:macroscopic_widths} a) is the evolution with $\epsilon$ in the resulting widths (FWHM) along $\phi$ and $\chi$ when fitting a Gaussian to the polefigures. Shown in Fig.~\ref{fig:macroscopic_widths} is the corresponding evolution of the local peak broadening. Apart from an offset at zero - reflecting that the pristine sample was not perfect - all parameters appear to be nearly linear in the applied strain. Moreover, we see that the local rotations (subfigure b)  on average are approximately 1/3 of the global ones (subfigure a).

\subsection{The pristine sample and the initial deformation}
\label{sub-pristine}

\begin{figure}[hbt!]
    \centering
    \includegraphics[width=0.8\linewidth]{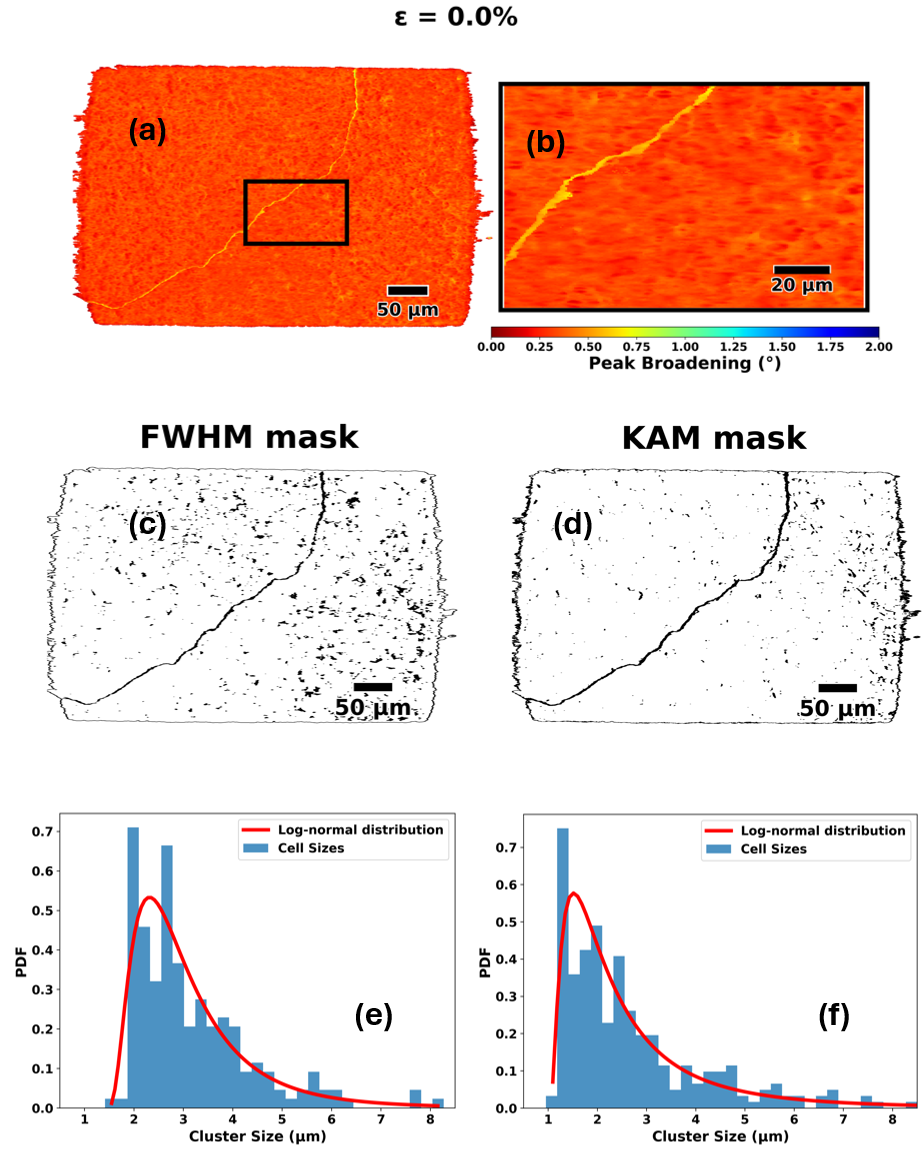}
    \caption{Properties of the pristine sample. a) the peak broadening map in false colors (the associated color bar is in degrees). b)  Zoom in on region marked by a box in a). c) Segmentation of peak broadening map for full field-of-view, with a threshold of 0.195 deg. d)  Corresponding KAM filter with a threshold of 0.01 deg. e) Histogram of cluster sizes in c) with a superposed best fit to a log-normal function. f) Histogram of cluster sizes in d) with a superposed best fit to a log-normal function.}
    \label{fig:pristine}
\end{figure}

The pristine sample is a bi-crystal with a low-angle boundary exhibiting a mis-orientation of 0.15 deg, cf. the pole figure, Fig.~\ref{fig:polefigures}, and the orientation map in Supplementary Video 1. This low angle boundary is clearly evident in the peak broadening map, cf. Fig.~\ref{fig:pristine} a). 
The dislocation content appears approximately constant at a level of 0.32 deg, about twice the misorientation angle. We interpret the excess angle as indicative of an elastic strain in the boundary of $\sim 0.005$. The spatial width of the boundary is in some places defined by the resolution, at others it is extended by up to 1 $\mu$m, cf. Fig.~\ref{fig:pristine} b).

Shown in Fig.~\ref{fig:pristine} c) is the result of applying a threshold to the $\Delta q$ map.  A corresponding Kernel-Average-Misorientation (KAM) mask is shown in  Fig.~\ref{fig:pristine} d). In both cases, the resulting "dislocation clusters" appear isolated and approximately randomly distributed (with the exception of the low angle boundary). The individual clusters differs substantially in shape and size between in the two images: we attribute this to the difference in dislocation population and strain components and noise. In Fig.~\ref{fig:pristine} e) and f) we have quantified the cluster size distributions using ImageJ. Given the populations size, both histograms are well described by log normal distributions.

The clusters as defined from thresholds on $\Delta q$ or  KAM are somewhat arbitrary, as they depend on the thresholds. As a complementary way to describe the dislocation ensemble, we can set a threshold on the weak beam image of the kind shown in Fig.~\ref{fig:overview} b). A histogram of the resulting size-distribution is again well described by a long-normal distribution.

Hence, three different analysis approaches all support the conclusion that the pristine sample comprise a set of dilute, non-interacting dislocation clusters, which are distributed randomly but homogeneously over the Field-of-View and with a size distribution which is consistent with a log-normal distribution.

\subsection{Long range order: autocorrelation based analysis}
\label{sec-autocorrelation}

 Using an autocorrelation function to determine order is at the root of diffraction and crystallography \cite{Als-Nielsen2011}. Following our previous work on an \emph{ex situ} sample \cite{Zelenika2024} we here use it to determine order on the length scale of micrometers. The script for the 2D autocorrelation itself is the same as used in the previous work; essentially it is based on use of the MATLAB function xcorr2. 

\begin{figure}[hbt!]
    \centering
    \includegraphics[width=0.7\linewidth]{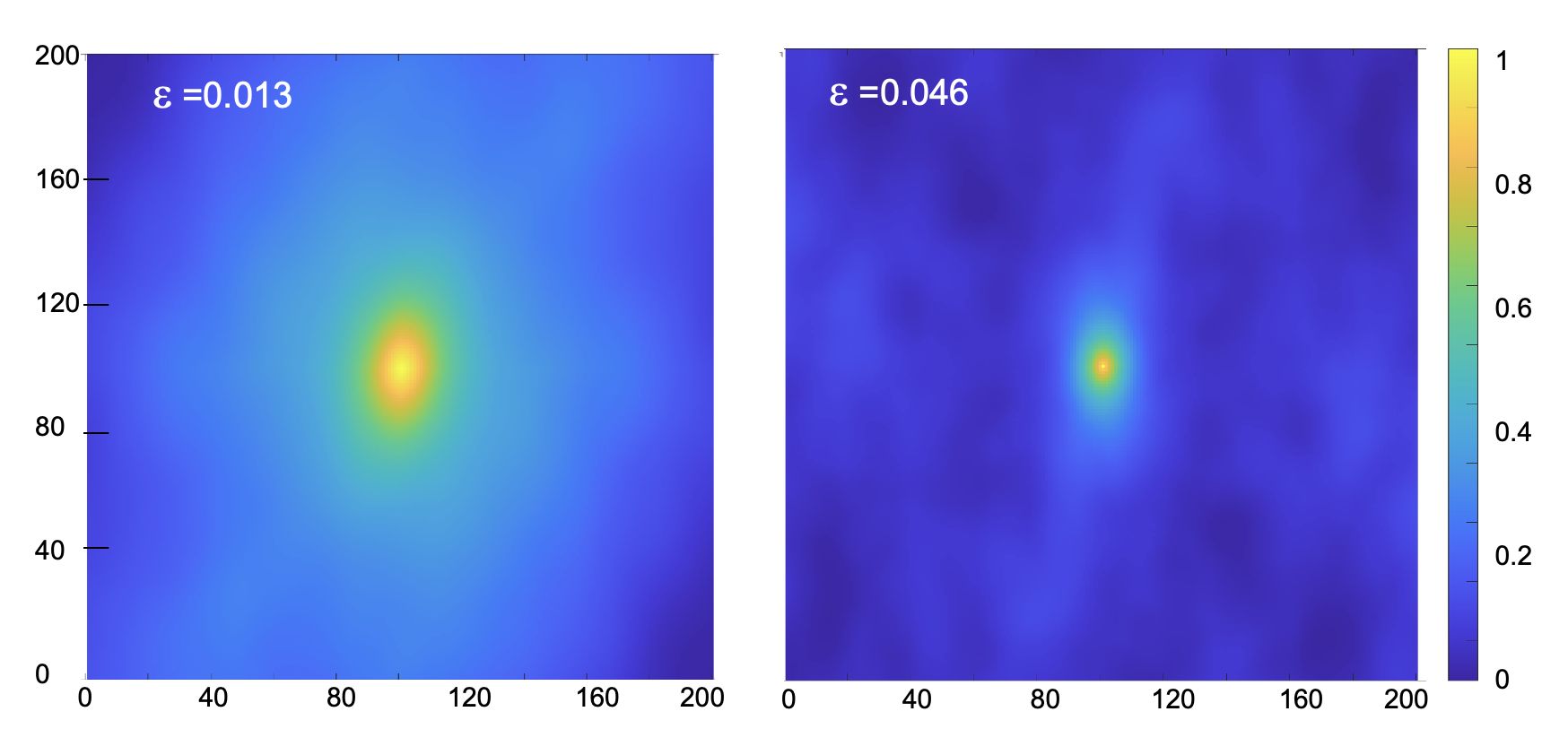}
    \caption{Auto correlation function for two applied strain levels}
    \label{fig:autocorr_in_2D}
\end{figure}

 The autocorrelation function for the undeformed and the most deformed states are shown in Fig.~\ref{fig:autocorr_in_2D}. The projections along the center lines are shown for all applied strains steps in Supplementary Video 3. The width (FWHM) of the central peak in the auto-correlation function along $x_l$ and $y_l$ are defined as the two "coherence lengths", 2$\xi_x$ and 2$\xi_y$, respectively. The resulting widths are tabulated in Table~\ref{tab:autocorrelation}.

\begin{table}[hbt!]
\centering 
\renewcommand{\arraystretch}{1.3}
\begin{tabular}{|c| c| c| c| c| c| c |}
    \hline 
    & 0 \%  & 0.8 \% & 1.3 \%  &   2.4 \% & 3.5 \% & 4.6 \%\\ 
    2$\xi_x$ & 31.2  &  24  & 12.1 & 6.2 & 4.2 & 3.6  \\ 
    2$\xi_y$ & 10.2  & 9.1  & 6.1  & 3.3 & 2.5 & 1.7  \\ 
    \hline 
\end{tabular}
\caption{Widths (FWHM) of the 2D autocorrelation function in directions $x_{\ell}$ and $y_{\ell}$: 2$\xi_x$ and 2$\xi_y$, respectively. They are listed in units of $\mu$m.}
\label{tab:autocorrelation}
\end{table}

In contrast to the \emph{ex situ} DFXM study on a similar sample but of a different orientation\cite{Zelenika2024}, there is here no evidence of long-range order emerging from the autocorrelation function. (The one exception are the side-loops along the $y_l$-direction for $\epsilon$ = 0 and 0.005 - these are due to the bicrystal nature of the pristine sample.) 

The  continuous decrease in the coherence lengths is attributed to the cell formation process. We note an anisotropy, with the coherent length in the $x_{\ell}$-direction consistently being 2-3 times larger than the one in the $y_{\ell}$-direction.  

\begin{figure}[hbt!]
    \centering
    \includegraphics[width=0.9\linewidth]{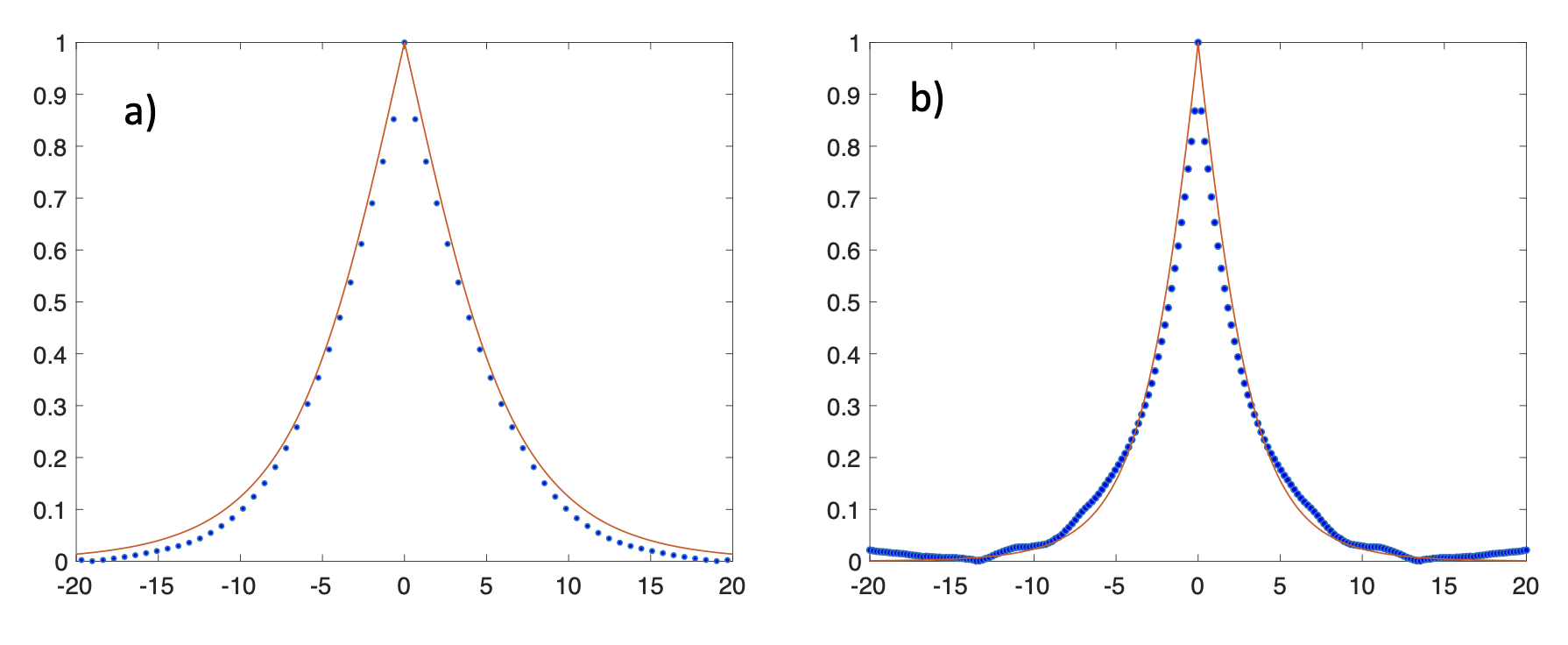}
    \caption{Comparison between experimental data for the autocorrelation (blue dots) and a hard sphere model with a log-normal size distribution for $\epsilon=0.046$. a) In direction $x_{\ell}$,  b) in direction $y_{\ell}$. In both cases the x-axis is in $\mu$m. }
    \label{fig:auto_comparison_46}
\end{figure}

In Fig.\ref{fig:auto_comparison_46} we present the autocorrelation function in directions $x_{\ell}$ and $y_{\ell}$ for $\epsilon = 0.046$. Superposed are the autocorrelation of a hard sphere model with non-interacting ellipses with a size distribution given by the anisotropic log-normal distribution found in section \ref{sec:cell_variable}. In order to compare the model and experimental data stereology has to be taken into account.  The autocorrelation function measures cord lengths. As such the width, $2\xi$ is related to the area, A, by $2\xi = \frac{4}{\pi} r = \frac{4}{\pi \sqrt{\pi}} \sqrt{A}$. On the other hand, the size, s, determined from the KAM is defined as $s = \sqrt{A}$. Hence, the hard sphere model is scaled by a factor $\frac{4}{\pi \sqrt{\pi}}$.  Given the fact that there are no free parameters in this comparison, this is an excellent correspondence, consistent with the model that  cells are fully formed at these points in time.  The autocorrelation profiles for $\epsilon = 0.035$ and $\epsilon = 0.024$ both exhibit a similar shape, but with decreasing applied strain the correlation lengths become larger and increasingly different from the cell size, consistent with the cell formation not being complete. 

At $\epsilon \le 0.013$ the autocorrelation function   appears to be a superposition of two functions, as shown in Supplementary Video 3. We interpret these data as evidence for a "two-phase" system. Part of the sample is in the process forming cells. Another part is still "undeformed" and this part consequently exhibits a longer correlation length.

\subsection{Cell analysis based on a variable KAM threshold.}
\label{sec:cell_variable}

EBSD based analysis typically defines grains/cells by means of a mask derived from a map of the Kernel-average misorientation, KAM \cite{Lassen1992, Kunze1993}.  Following this tradition we will in the following define cells based on a threshold for the misorientation $  \theta_{\mathrm{mis}} $ and an isotropic kernel of size 2.

\begin{figure}[hbt!]
    \centering
    \includegraphics[width =\linewidth]{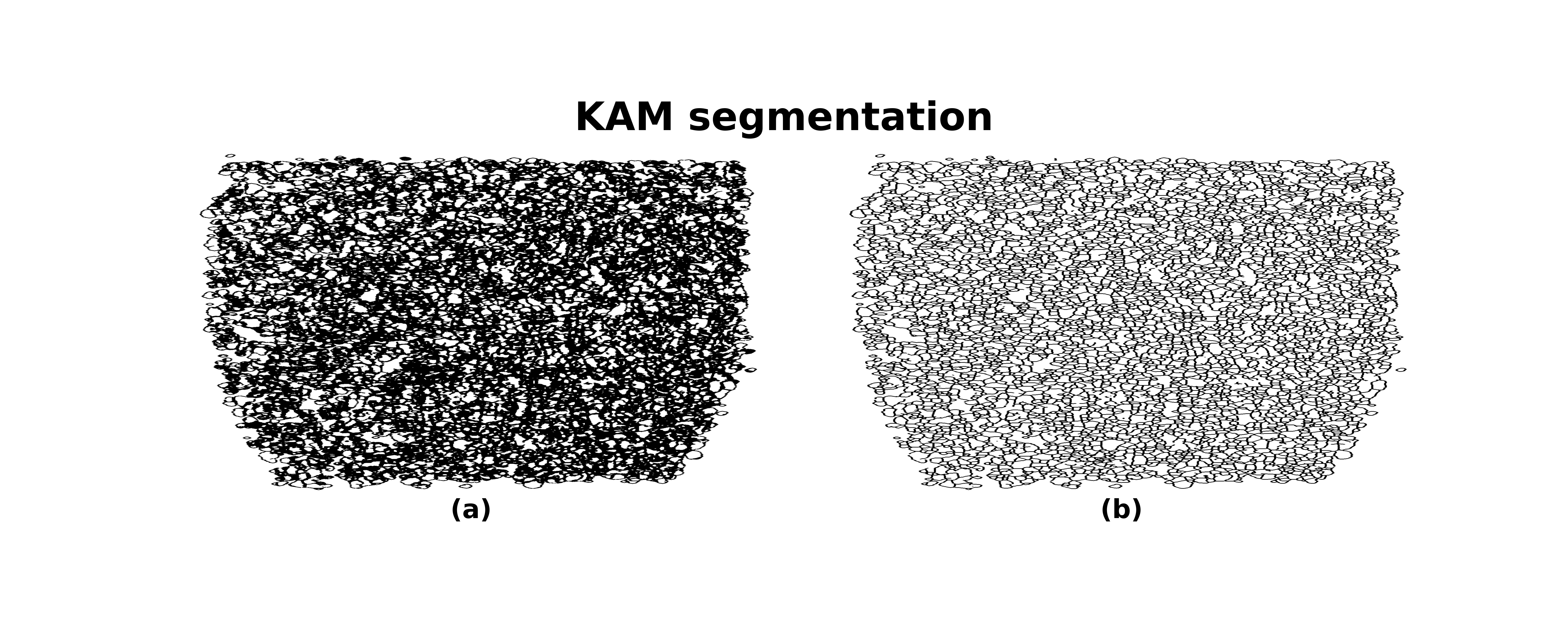}
    \caption{Example of KAM mask for $\epsilon = 0.046$. a) Segmentation based on threshold of 0.034 deg in misorientation. b) corresponding KAM mask after morphological operations. }
    \label{fig:KAM_mask_details}
\end{figure}
As we in DFXM with one reflection only map two orientation degrees of freedom, the definition for misorientation is revised to be
\begin{align}
    \theta_{\mathrm{mis}} & =  \sqrt{ (\Delta \phi)^2 + (\Delta \chi)^2   }.
\end{align}

Shown in Fig.~\ref{fig:KAM_mask_details} a) is a segmented image for $\epsilon = 0.046$. The fraction of material below the threshold is in this case 71 \%. Next morphological operations are applied to this binary image. Specifically, this image is skeletonized followed by a dilation of 1,  implying that all cell boundaries have a thickness of 3 pixels. For the analysis in general the boundary thickness is 1 pixel. The resulting KAM mask is shown in Fig.~\ref{fig:KAM_mask_details} b). This is overlaid on the orientation map in the last frame in Supplementary Video 4.

Based on an ansatz of a linear scaling with the applied strain we apply a variable KAM threshold that increases with the applied strain. Specifically, we keep the fraction of the segmented image above the threshold constant: $ \theta_{\mathrm{mis}} = \sim 71 $\%.  This corresponds to thresholds of 0.10, 0.20, 0.32 and 0.41 deg for $\epsilon = 0.013, 0.024, 0.035$ and $0.046$ deg, respectively. The almost linear correlation is remarkable and justifies the linear ansatz.

The KAM algorithm leads to the generation of a site-filling map of domains. For lower degrees of applied strain one or more of these domains will represent the matrix:  the undistorted part of the sample. Such domains (here defined by having areas larger than (25 $\mu$m$)^2$) are removed from the set. So are domains with a size smaller than 10 pixel units. Finally, due to the $(\phi,\chi)$ region scanned being a bit too small a minor fraction of the $\epsilon = 0.035 $ and 0.046 maps are filled with voids - again these regions have been removed. The remaining domains are identified as cells. The resulting cell maps are provided as Supplementary Video 1, with some of teh frames also reproduced in  Fig.~\ref{fig:cell_formation}.

The scipy function \todo{name} enables the generation of statistics of the structural parameters embedded in the cell map.  In addition, nearest neighbors are identified by dilating each cell slightly and detecting overlapping voxels. The  volume fraction and mean cell sizes are reporoduced as function of the applied strain in Fig.~\ref{fig:cell_formation}. To enable a direct comparison between load steps, the number of cells and area fractions have been calibrated to the same total area. Other key statistical parameters are provided in Table~\ref{tab:cell_stats}. 

\begin{table}[hbt!] 
\centering 
\renewcommand{\arraystretch}{1.3}
\begin{tabular}{|c| c| c| c| c| c| c| c| }
    \hline 
    $\epsilon$ & Number & Avg. size &  Avg. An-    & Avg. misorienta- \\ 
    &     cells    & ( in $\mu$m) & isotropy & tion (in \degree) \\
    \hline
    0.000 & 47 & 2.11 & 0.63  & NaN \\
    0.005 & 100 & 2.46  & 0.64  & NaN \\
    0.008 & 166 & 2.81  & 0.62  & NaN  \\
    0.013 & 4033 & 4.81  & 0.59  & 0.06  \\ 
    0.024 & 4125 & 5.02  & 0.58 & 0.14  \\ 
    0.035 & 4383 & 4.87  & 0.601 & 0.25 \\ 
    0.046 & 4090 & 4.59  & 0.61 & 0.32 \\ 
    \hline 
\end{tabular}
\caption{Statistics of the cells generated by the KAM filter for a variable threshold of $\theta_{\mathrm{KAM}}$. The size is defined as the square-root of the area. The anisotropy is defined as the ratio between the major axis and the minor axis of the cells.}
\label{tab:cell_stats}
\end{table}

\subsubsection{Cell shape anisotropy}
\begin{figure}[hbt!]
    \centering
    \subfloat[]{\includegraphics[width=0.35\linewidth]{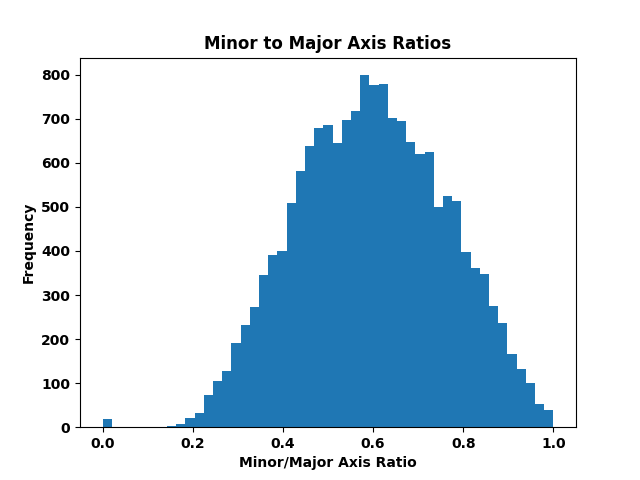}}
    \hspace{0.5cm}
    \subfloat[]{\includegraphics[width=0.35\linewidth]{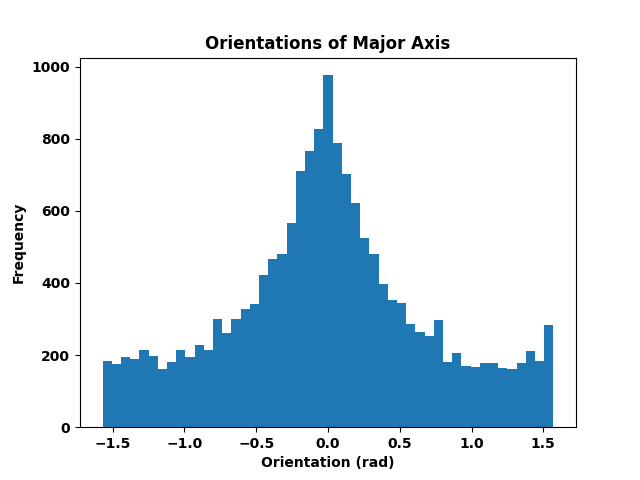}}
    \caption{Cell statistics on aspect ratios of the cells for $\epsilon=0.046$. The 0 orientation in b) corresponds to the direction of the $x_l$-axis.  }
    \label{fig:cell_aspect_ratio}
\end{figure}

Shape information arising from statistics on the $\epsilon = 0.046$ set of cells is summarised in Fig.\ref{fig:cell_aspect_ratio}. The cells are on average elongated in the $x_l$-directions, consistent with the result of the auto-correlation analysis.

\subsubsection{Cell mis-orientations}

The local orientation relationships for the middle layer at applied strain $\epsilon = 0.046$ have been characterised in two ways: 

\begin{figure}[hbt!]
    \centering
     \includegraphics[width = 0.7\linewidth]{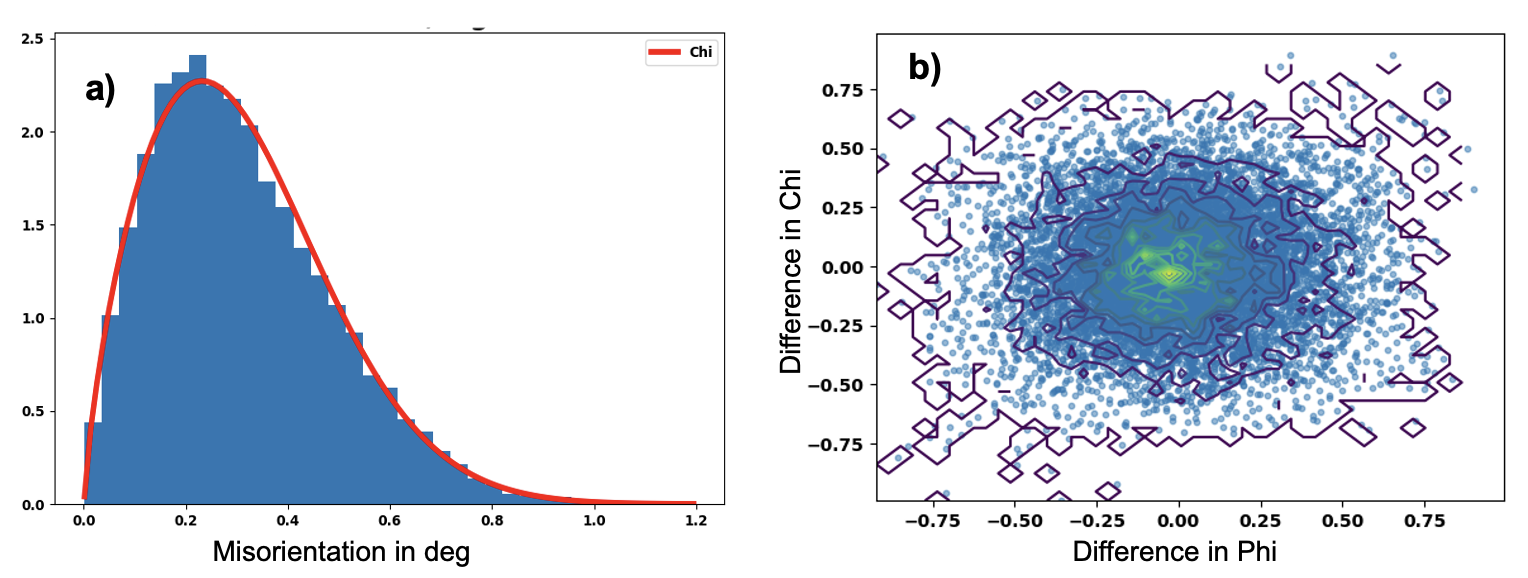}
    \caption{Orientation correlation between neighboring cells for $\epsilon=0.046$, I. a) Misorientation angle with best fit to a chi-function superposed. b) the misorientation axis. The contour line intensities are decreasing from (0,0) and outwards with a constant step in intensity.}
    \label{fig:misorientation_analysis}
\end{figure}

    \emph{Firstly}, the misorientation angle and misorientation direction between neighboring cells is deduced from the set of cells generated above. The resulting mis-orientation angle and axis distributions are shown in Fig.~\ref{fig:misorientation_analysis}.   The sample appears to be isotropic to a high degree of accuracy.   

\begin{figure}[hbt!]
    \centering
    \includegraphics[width=0.9\linewidth]{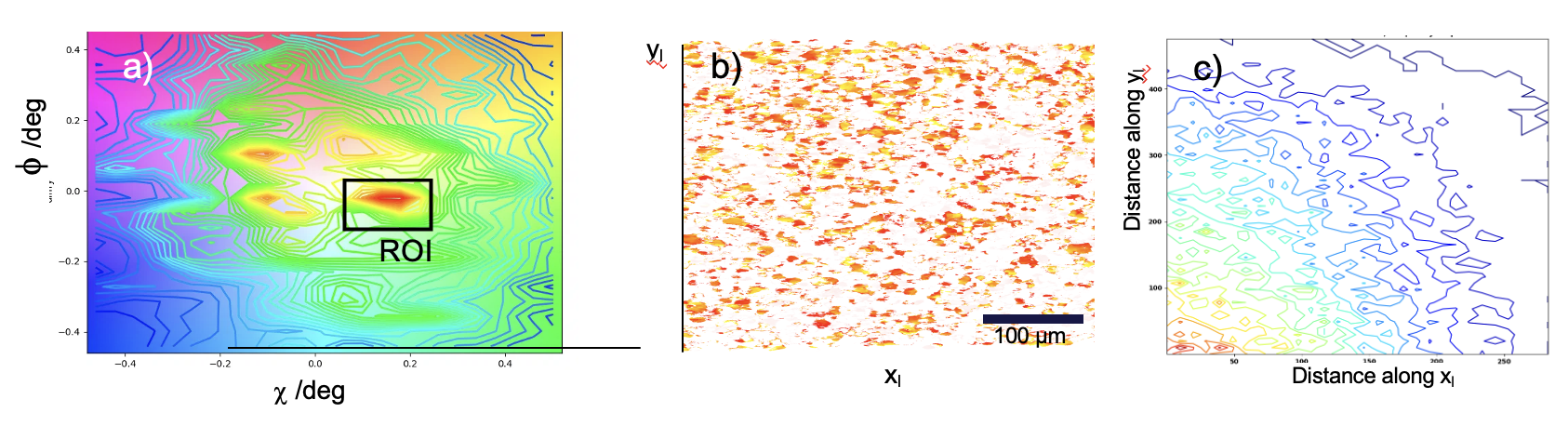}
    \caption{ Orientation correlation for $\epsilon=0.046$, II. a)  Polefigure composed of the COM positions with region-of-interest, ROI, superposed. b) Zoom in on cell map, showing only cells with an orientation within the ROI. c) A distance frequency map based on the set of cells identified in b). The contour line intensities are decreasing from (0,0) and outwards.  }
    \label{fig:anisotropy_ROI}
\end{figure}

\emph{Secondly}, as shown in Fig.~\ref{fig:anisotropy_ROI} a) we define a region-of-interest, ROI, around one of the poles in the pole-figure. The area of the ROI  is $ 0.1 \deg \times 0.2 \deg$. The subset of cells with orientations within the ROI, $S_{ROI}$ are shown in Fig.~\ref{fig:anisotropy_ROI} b). Next distances $d_{ij}$ between center positions $((x_l)_i, (y_l)_i)$ of all cells   $ i,j \in S_{ROI}$ were calculated. The angular distribution of $d_{ij}$ is shown in Fig.~\ref{fig:anisotropy_ROI} c). Again within experimental error the data are completely isotropic. 

\subsubsection{Cell size distribution - comparison with theory}

\begin{figure}[hbt!]
    \centering
    \includegraphics[width=0.95\linewidth]{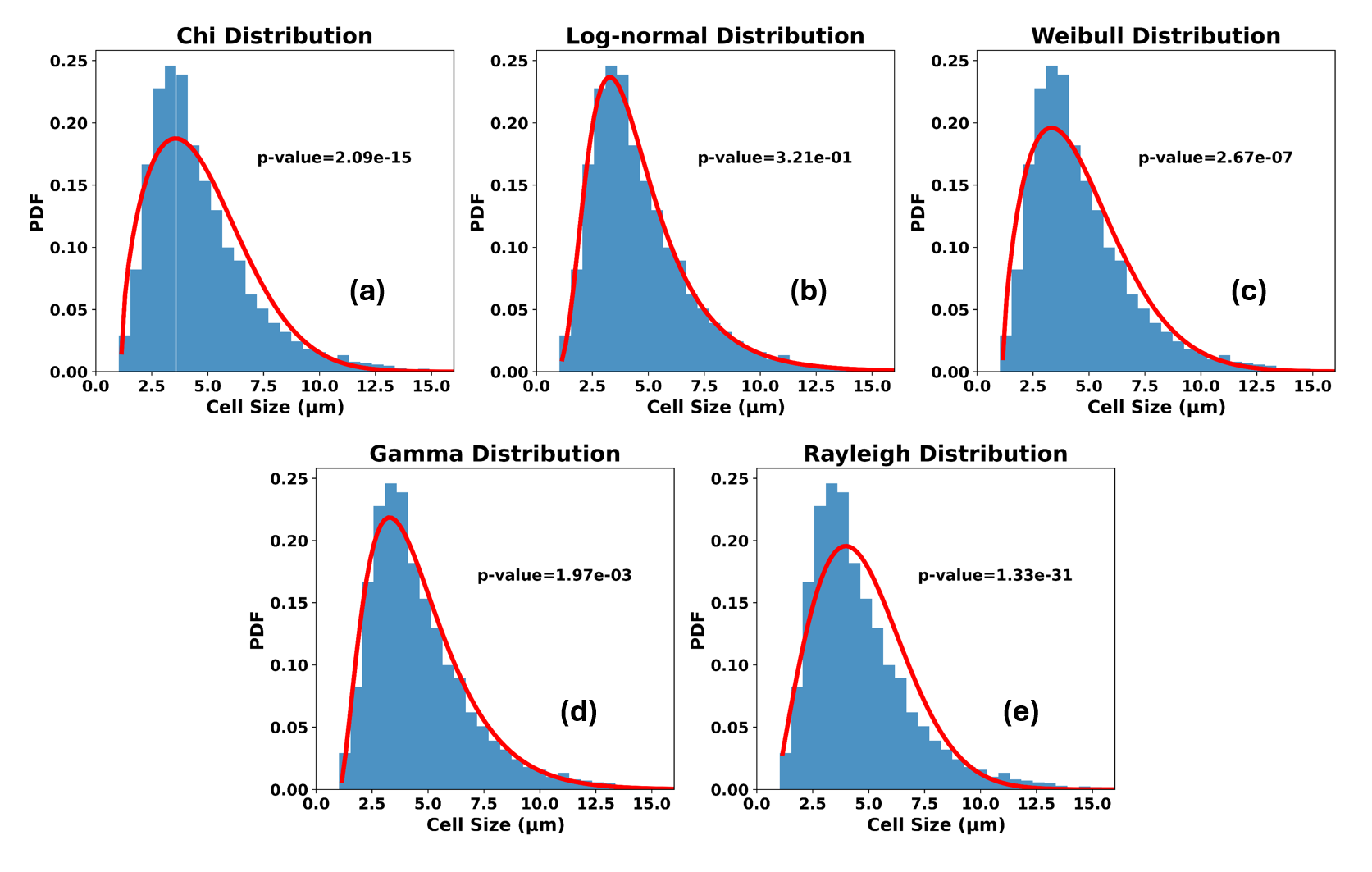}
    \caption{Test of models for cell size distribution for $\epsilon = 0.046$. Superposed (red lines) are the best fits to model functions. Listed are also the results of Kolmogorov Smirnov tests.}
    \label{fig:grainsize_testmodels}
\end{figure}
 A survey of the compatibility of a set of model distribution functions with the cell size distribution generated by the KAM filter for $\epsilon = 0.046$ is shown in Fig.~\ref{fig:grainsize_testmodels}. Visual inspection clearly identify the log-normal distribution as superior. A non-parametric test, the Kolmogorov-Smirnov, was used to quantify the fits. The $p$ values listed in the legends in Fig.~\ref{fig:grainsize_testmodels} represent the likelihood that the experimental data can be represented by the various distributions. With $p=0.57$ the data are clearly consistent with the log-normal distribution, which is remarkable given the large population. On the other hand, with the usual threshold for statistical significance ($p_{\mathrm{limit}} = 0.05$) none of the other distributions pass the test. The fitting and tests were performed by standard Python SciPy code.

\begin{figure}[hbt!]
    \centering
    \includegraphics[width = 0.9\textwidth]{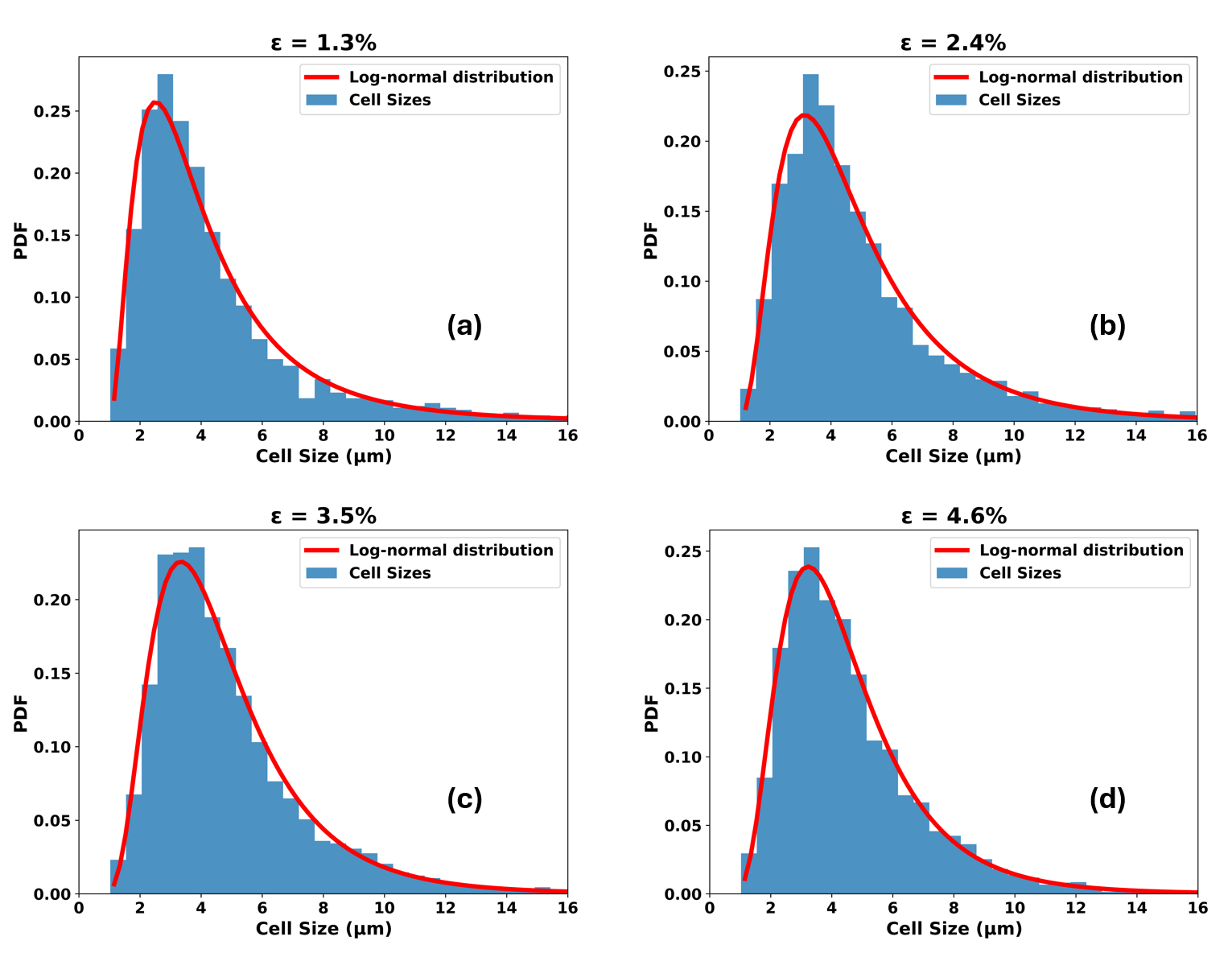}
    \caption{Log-normal fits to cell size distribution as function of applied strain, $\epsilon$ with variable misorientation threshold.}
    \label{fig:lognormal_fits}
\end{figure}

Remarkably, the log-normal distribution is a good approximation in the entire range where applying a KAM filter is meaningful: for $\epsilon = 0.013$ to $\epsilon = 0.046$, cf. Fig.~\ref{fig:lognormal_fits}. The resulting optimised values $\mu$ and $\sigma$ are listed in Table~\ref{tab:log_normal} as well the mean (by definition equal to $\exp(\mu + \sigma^2/2)$. Also shown is the p value for the corresponding test. 
\begin{table}[hbt!]
\centering 
\renewcommand{\arraystretch}{1.3}
\begin{tabular}{|c| c| c| c| c| }
    \hline 
     $\epsilon$  &  0.013 &  0.024 & 0.035 & 0.046 \\ 
     \hline 
    $\mu$ (in $\mu$m) & 1.26  & 1.27 & 1.34 & 1.32  \\ 
    $\sigma$ (in $\mu$m) & 0.60 & 0.62 & 0.53 & 0.51\\ 
    mean (in $\mu$m) & 4.81 & 5.02 & 4.87 & 4.59 \\ 
    $p$ & 0.034 & 0.052  & 0.183  & 0.321 \\
    \hline 
\end{tabular}
\caption{Cell size distribution as defined by variable threshold $\theta_{mis}$. Fitted values for the two parameters in the log-normal distribution: $\mu$ and $\sigma$. Also listed are the p-values from corresponding Kolmogorov Smirnov tests.}
\label{tab:log_normal}
\end{table}

\subsubsection{Cells analysis with fixed KAM threshold}
\label{sec:cells_fixed_threshold}

\begin{figure}[hbt!]
    \centering
    \includegraphics[width=0.4\linewidth]{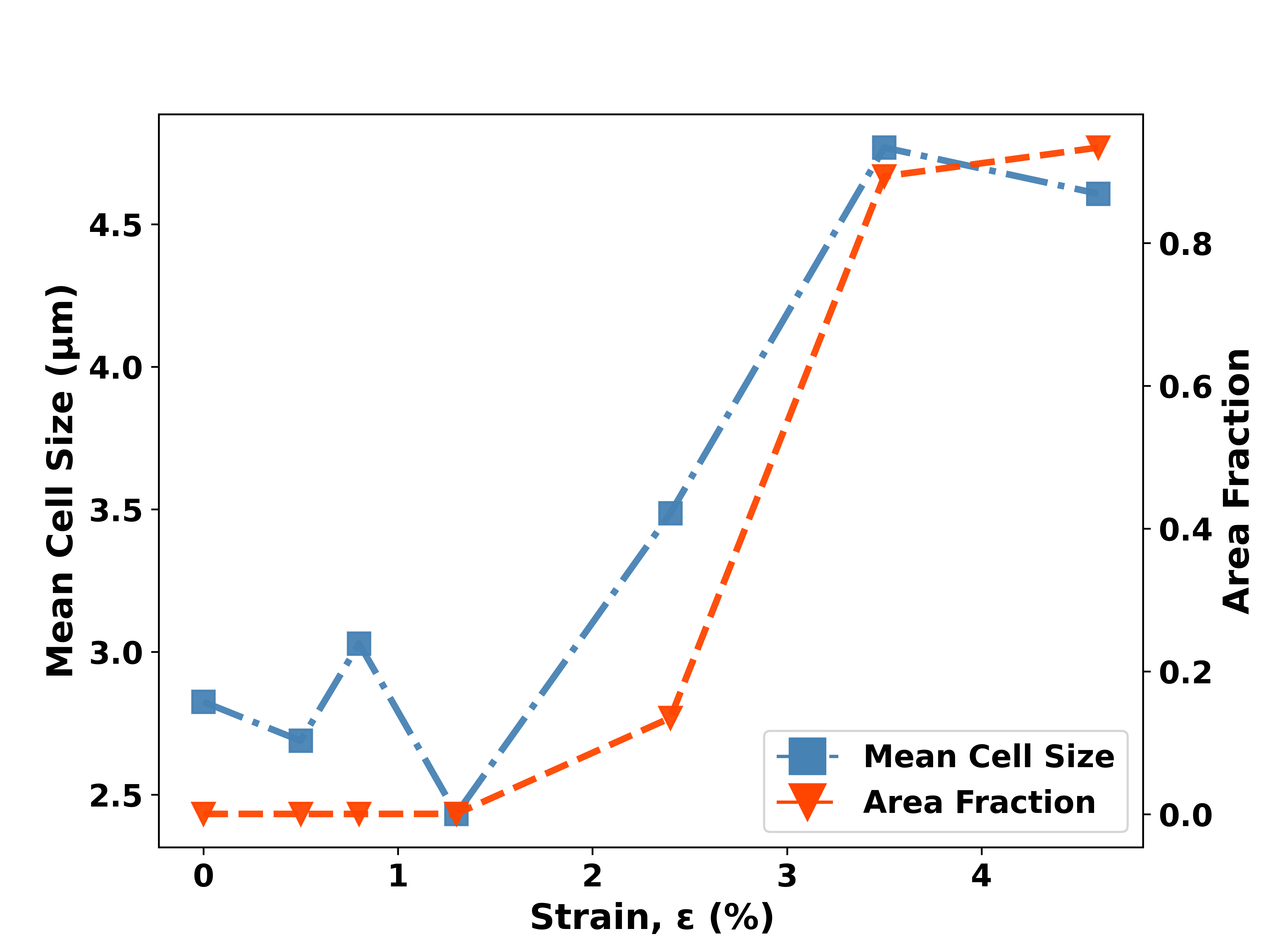}
    \caption{Area fraction of cells and average cell size as function of $\epsilon$ for a fixed misorientation threshold.}
    \label{fig:size_fixed_KAM}
\end{figure}

\begin{figure}[hbt!]
    \centering
    \includegraphics[width = 0.9\textwidth]{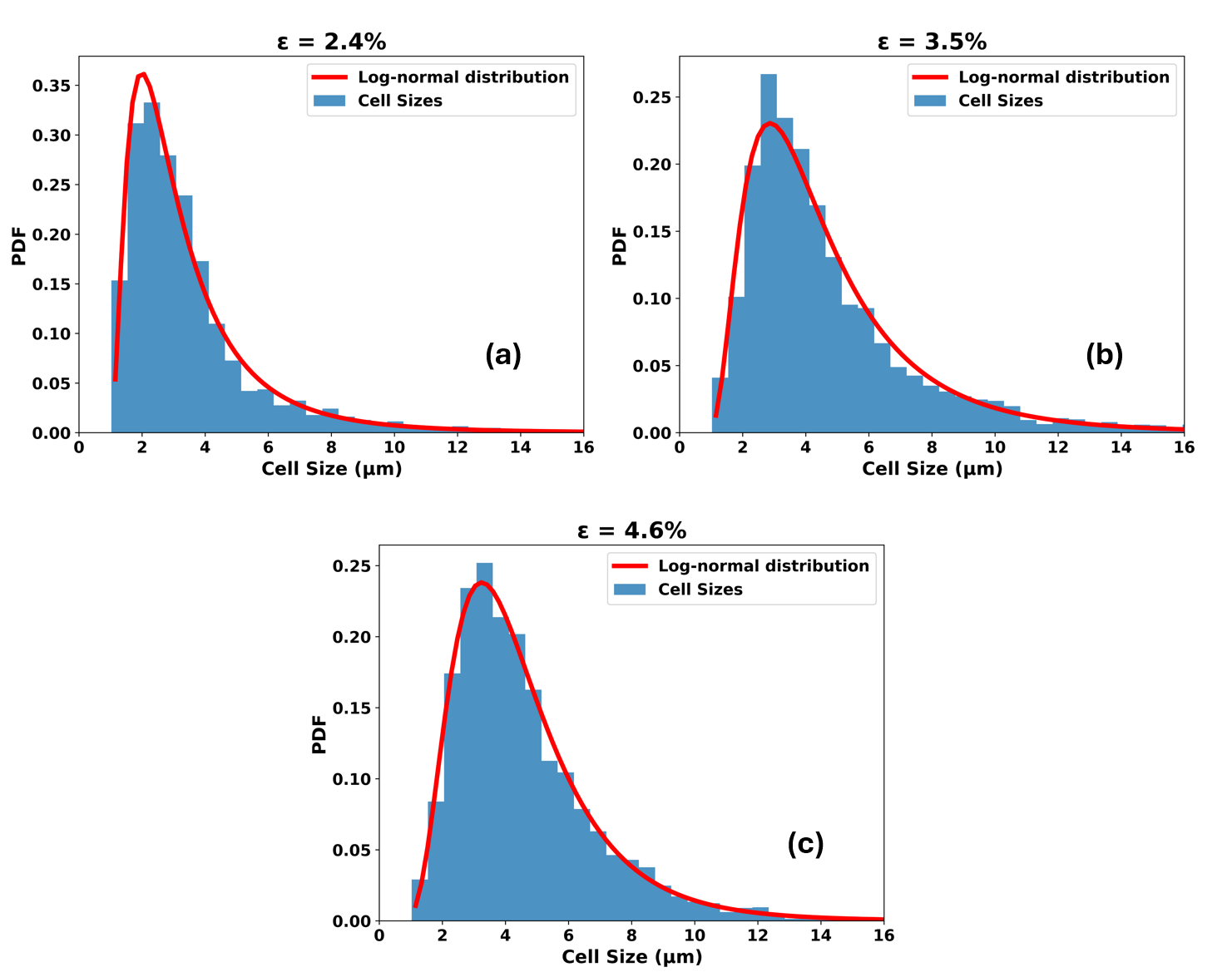}
    \caption{Cell size distributions as function of applied strain for a fixed misorientation threshold $\epsilon$. Superposed are log-normal fits. For direct comparison with Fig.~\ref{fig:lognormal_fits}. }
    \label{fig:fixed_lognorm}
\end{figure}

For completion we provide an analysis similar to that of section~\ref{sec:cell_variable} for the case of keeping the KAM threshold fixed at the value optimised for $\epsilon = 0.046$:  $  \theta_{\mathrm{mis}}  = 0.034 \deg$. Results for volume fraction of cells, mean size and cell size distribution are shown in Fig.~\ref{fig:size_fixed_KAM} and \ref{fig:fixed_lognorm}, respectively.
The conclusions in relation to the size distribution are identical to those obtained with a variable threshold, except for the cells forming later, as is to be expected.

\begin{table}[hbt!]
\centering 
\renewcommand{\arraystretch}{1.3}
\begin{tabular}{|c| c| c| c| c| }
    \hline 
     $\epsilon$  &  0.013 &  0.024 & 0.035 & 0.046 \\ 
     \hline 
    $\mu \; (\mathrm{in} \; \mu$m) & NaN  & 0.66 & 1.19 & 1.36 \\ 
    $\sigma \; (\mathrm{in}  \; \mu$m) & NaN & 0.76 & 0.65 & 0.51 \\ 
    $\mathrm{mean} \; (\mathrm{in}  \; \mu$m) & NaN & 3.49 & 4.77 & 4.59 \\ 
      p & NaN & 0.219  & 0.036  & 0.678 \\
    \hline 
\end{tabular}
\caption{Cell size distribution as defined by fixed threshold $\theta_{mis}$. Fitted values for the two parameters in log-normal distribution: $\mu$ and $\sigma$. Also listed are the p values from corresponding Kolmogorov Smirnov tests.}
\label{tab:log_normal_fixed_threshold}
\end{table}

\subsection{Dislocation density distribution }

\begin{figure}[hbt!]
    \centering
    \includegraphics[width = 0.9\textwidth]{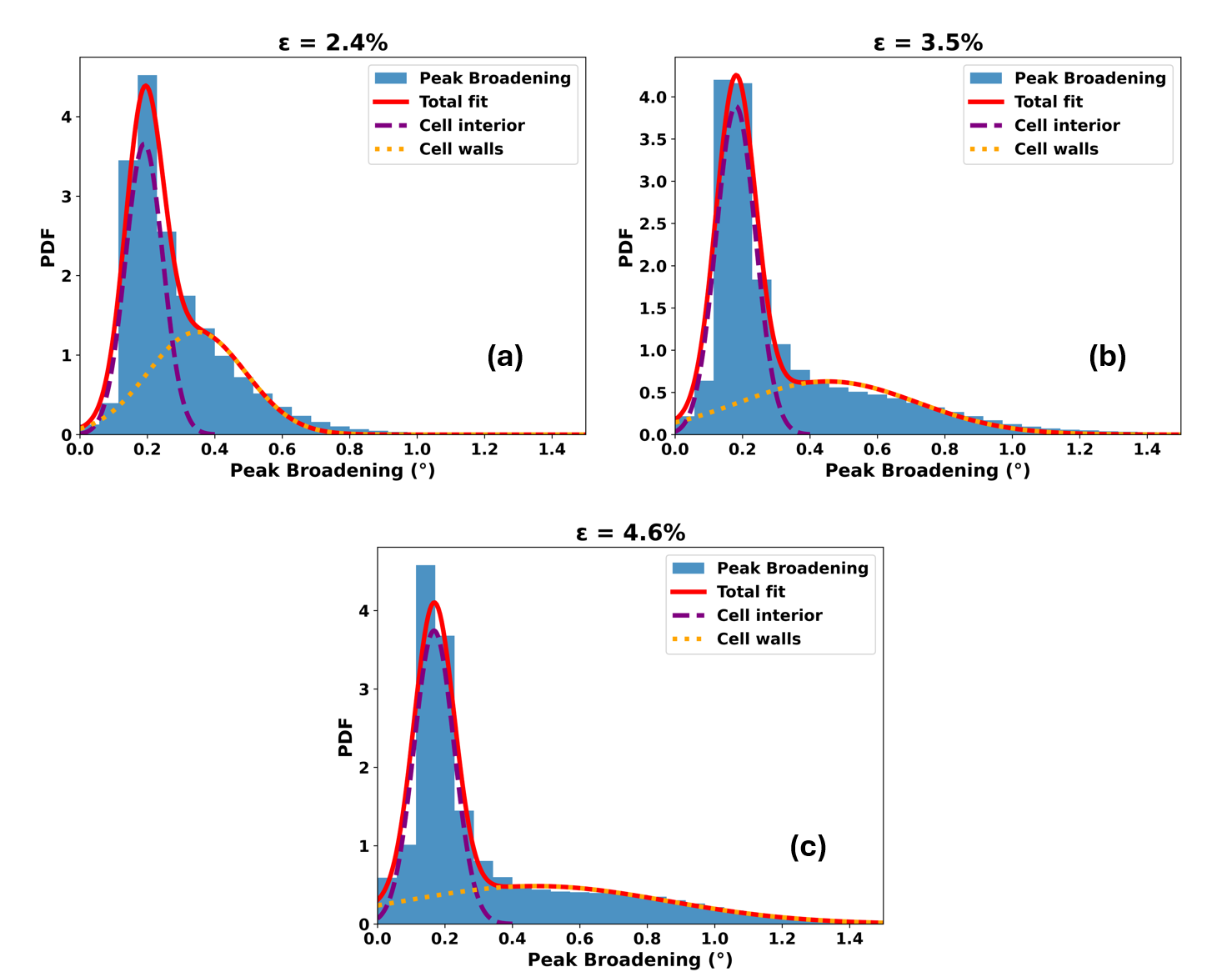}
    \caption{Distributions of peak broadening $\Delta q$ as function of  $\epsilon$: a) 0.024, b) 0.035, c) 0.046. Superposed are best fits to a model comprising a bimodal Gaussian distribution (red curves). Also shown are the contributions to the model from the cell interiors (purple) and cell walls (yellow curves). }
    \label{fig:bimodal}
\end{figure}

In Fig.~\ref{fig:bimodal} we plot the distribution of the average local peak broadening n directions $\phi$ and $\chi$. As described in in Section~\ref{sub_theory_peakbroadening} this is a proxy for the square root of the total dislocation density. It appears that a fit to a bimodal Gaussian distribution is satisfactory for $\epsilon = 0.024$, 0.035 and 0.046. Following the modelling work by Chen {et al.} \cite{Chen2021} it is natural to identify the two peaks as arising from regions where the response is elastic and plastic, respectively. We note that the statistics is not sufficient to determine whether the plastic component is normal or log-normal in nature. For convenience we here used a normal distribution. Moreover, by inspection of the corresponding maps, see Supplementary video 2,  we infer that these two peaks approximately corresponds to nearly dislocation free regions and "stored dislocations regions": cell interiors and walls.  

\begin{table}[hbt!] 
\centering 
\renewcommand{\arraystretch}{1.3}
\begin{tabular}{|c| c| c| c| c|}
    \hline 
     $\epsilon$  & 0.024 & 0.035  &  0.046\\ 
     \hline
    Amplitude peak 1  & 3.30  & 3.58 & 3.43  \\ 
    Mean peak 1  & 0.19 & 0.18 & 0.17  \\ 
    sigma peak 1  & 0.04   & 0.04  & 0.04   \\
    Amplitude peak 2  & 1.40  & 0.67 & 0.51 \\ 
    Mean peak 2 &   0.35 & 0.45 & 0.45 \\ 
    sigma peak 2  & 0.10 & 0.19 & 0.29  \\
    Area ratio peak1/peak2  & 0.90  & 1.15 & 0.97 \\
    \hline 
\end{tabular}
\caption{Optimised parameters resulting from the bimodal Gaussian fits to the average peak width, cf. Fig.~\ref{fig:bimodal}. The units of spatial variables is $\mu$m. Also shown is the ratio between the areas of the two peaks. }
\label{tab:bimodal}
\end{table}

The resulting optimized parameters are provided in Table~\ref{tab:bimodal}. To first order the area fraction of the two components is constant.  As illustrated in Fig.~\ref{fig:evolution}
the average peak width of the dislocation rich component grows approximately linearly with the applied strain. 

\begin{figure}[hbt!]
    \centering
   \includegraphics[width=0.7\linewidth]{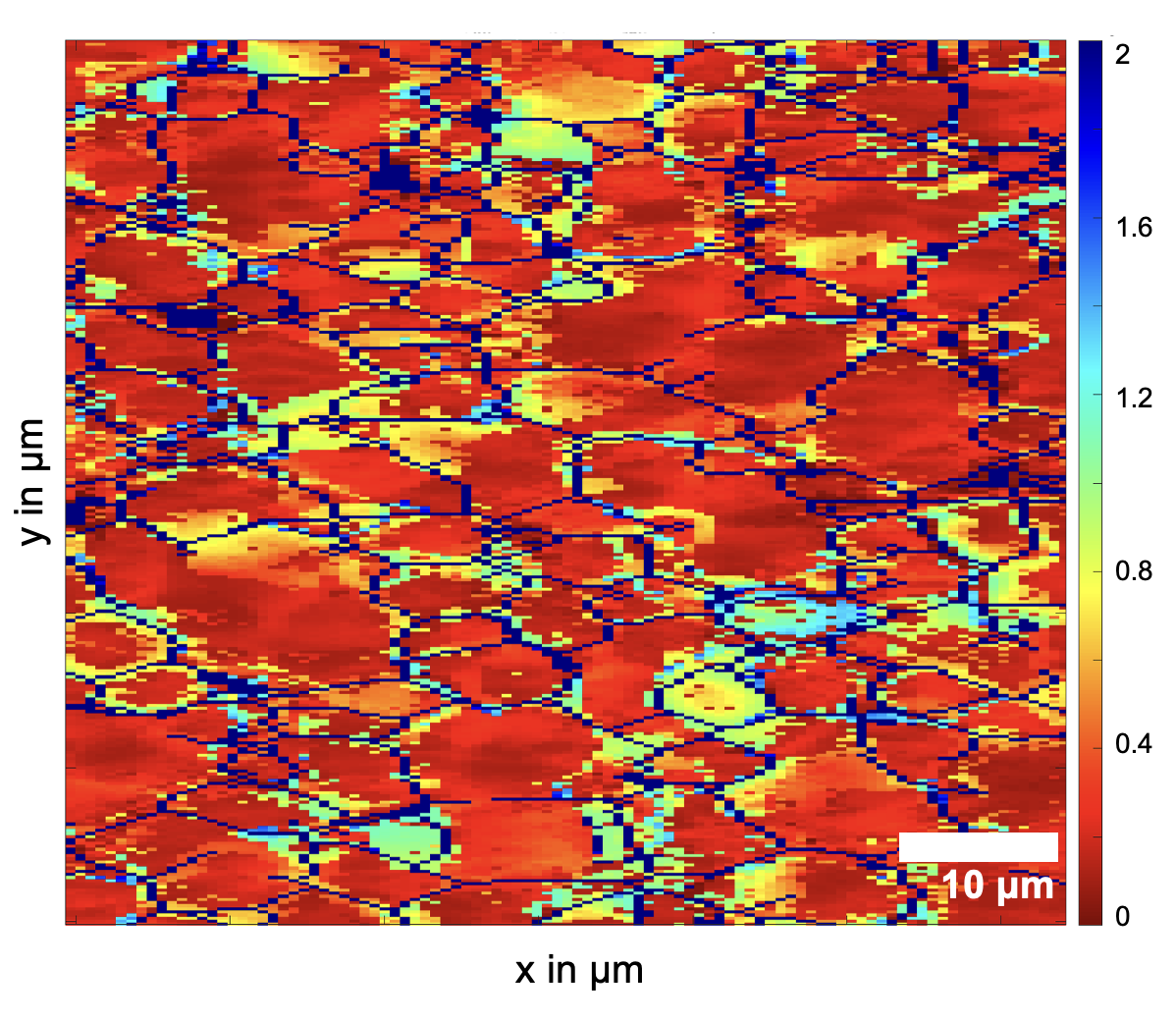}
    \caption{Region of interest of the peak broadening map for $\epsilon = 0.046$ with angular values in degrees as presented by the color bar. Superposed is the KAM mask (dark blue lines). } 
    \label{fig:fwhm_KAM_overlaid}
\end{figure}
\vspace{0.5cm}

Shown in Fig.~\ref{fig:fwhm_KAM_overlaid} is the spatial distribution of (the square root of) the total dislocation densities in relation to the cell boundaries, as defined by the KAM filter. Consistent with results from TEM the boundaries tend to decorate with dislocations within a boundary region of order 1 micrometer. in addition we identify some larger regions of relatively high density. Some cells have a high density throughout their area. Statistical tests shows that these are all relatively small cells. We attribute them as being the "bottom" or "top" of larger cells which have their center-of-mass outside the layer inspected.

\begin{figure}[hbt!]
    \centering
    \includegraphics[width=0.55\linewidth]{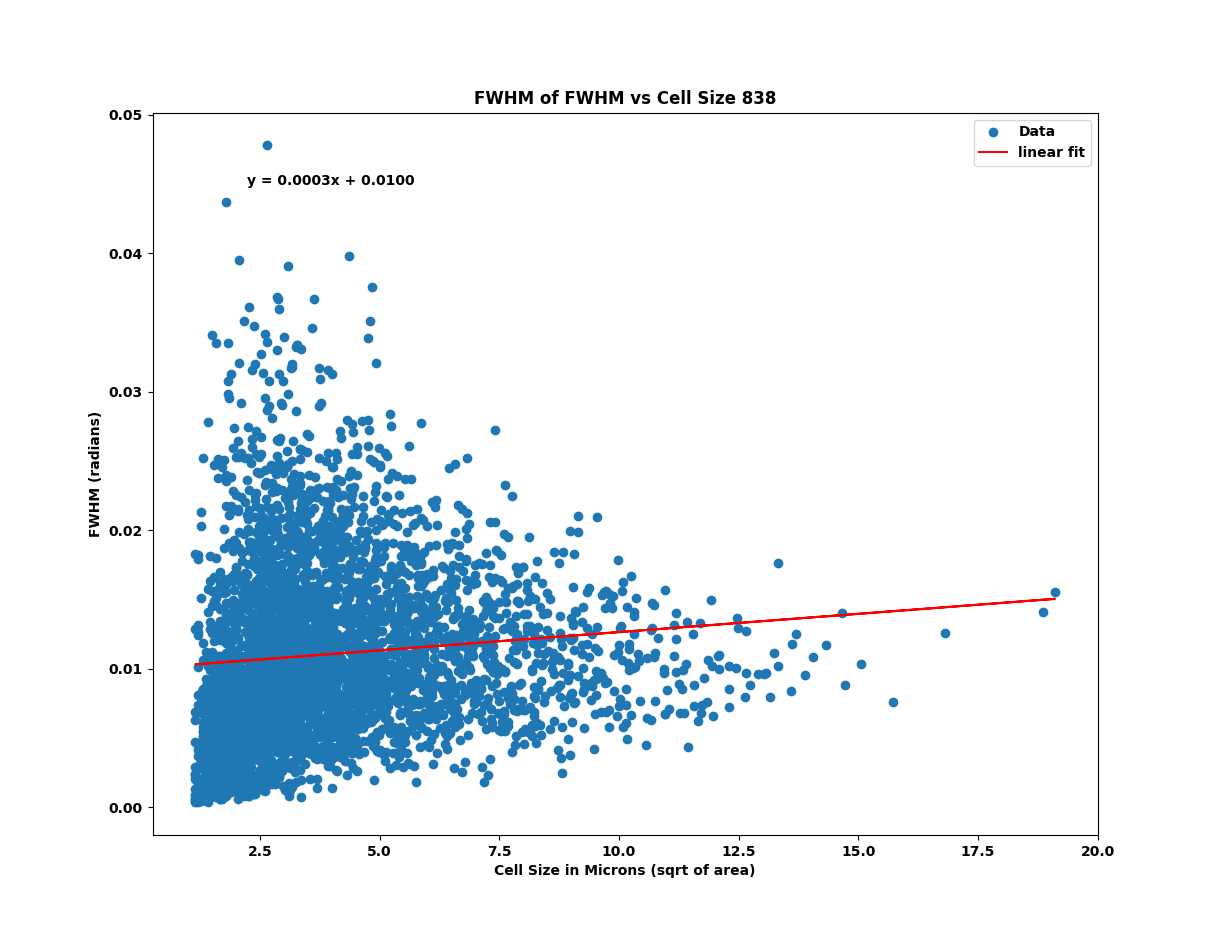}
    \caption{Correlation between cell sizes and the range of the peak broadening data, FWHM ($\Delta q^{\mathrm{cell}}$), within each cell for $\epsilon = 0.046$. Superposed is the best fit to a linear relationship (red line).}
    \label{fig:corr_strain_and_gradient_inside_cells}
\end{figure}

Inspection also reveals that while most of the cells have a uniformly low density inside, some are heterogeneous in the sense that there are one or more subdomains with significantly larger density.  Fig.~\ref{fig:corr_strain_and_gradient_inside_cells} confirms the existence of a positive correlation between such heterogeneity and cell size.  
Speculating that such high-density sub-domains are precursors to new cells, and that the likelihood for fragmentation on average is linear in area, this insight outlines a mechanism that conforms with the concept of multiplicative stochastic processes. This \emph{hypothesis} implies a birth process that is the opposite of the one in recrystallization, where the nuclei are less deformed than the matrix.

\subsection{Comparison with literature values for larger applied strains}
\label{sec:compare_literature}

To the knowledge of the authors there are no quantitative microstructural data for the strain range studied in this work. In the following we compare with the work of Huang, Hansen \emph{et al.} \cite{Huang1997, Hansen2001b}. They report on the TEM results of tensile strained polycrystalline 99.996 \% pure aluminium in the range $\epsilon = 0.05 $ to $\epsilon = 1.0 $. Three characteristics types of microstructures appear as function of grain orientation. For [111] they report on the so-called type 3 structure, a cell block structure where the dense dislocation walls are rotated by about 40 degrees to the TA. However, this characterisation was based on thin foil studies in a plane with a normal $\parallel$ TA. In contrast, TEM data acquired in the plane $\perp$ TA exhibited no discernible band structure \cite{Kawasaki1980}.  The current data are acquired in a plane that is rotated by only $\theta = 10 \; \deg$ from the plane with a normal $\perp$ TA.  The spatial anisotropy observed may be related to an out-of-plane ordering, but similar to TEM, otherwise there is no trace of band structure in the plane observed.  

\begin{figure}[hbt!]
    \centering
    \includegraphics[width=0.5\linewidth]{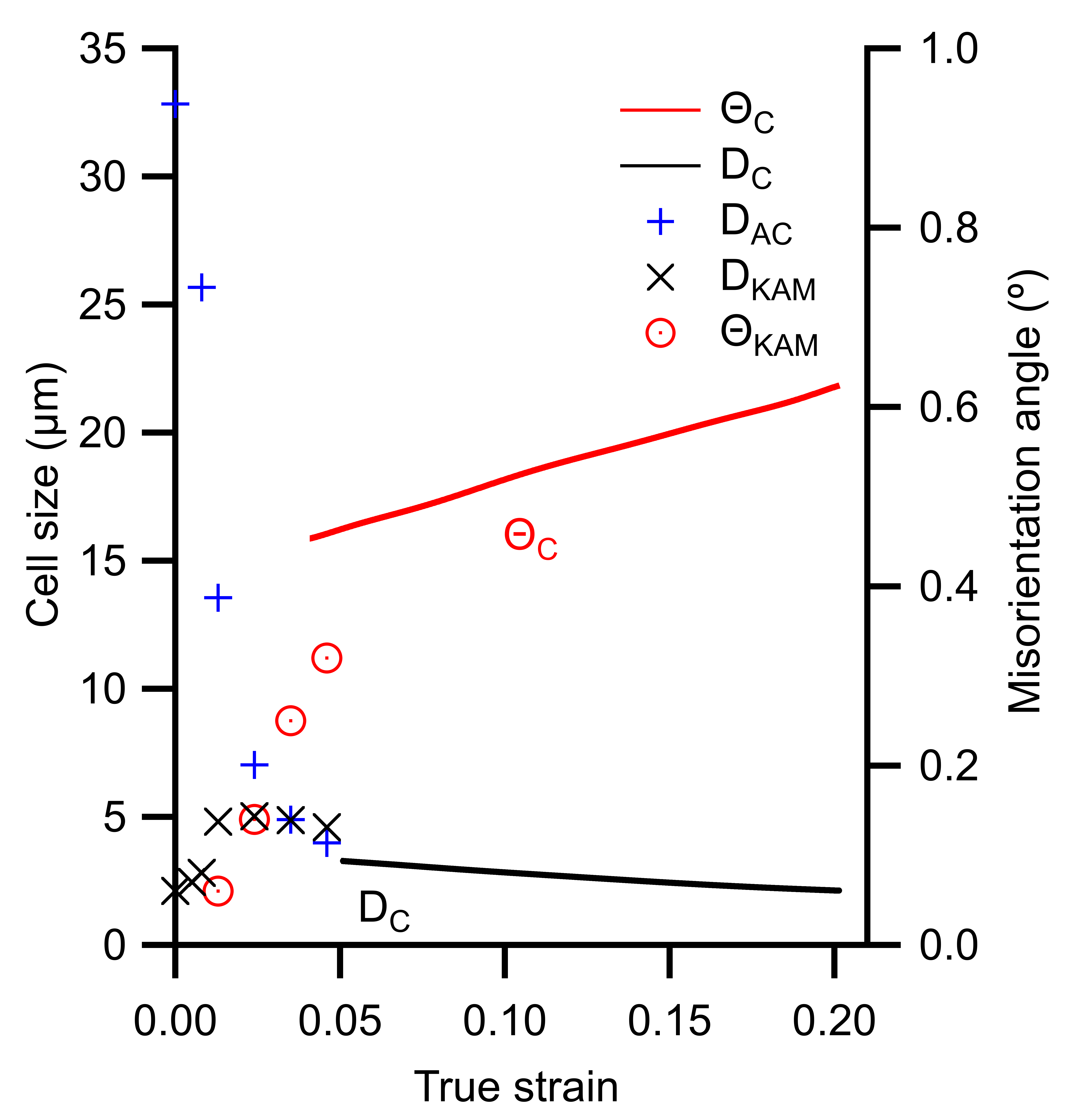}
    \caption{Comparison of the structural properties measured by DFXM (symbols) with literature values based on TEM \cite{Hansen2001} (lines). For the average cell size  the literature values  ($D_C$)  are compared to the results of the cell analysis listed in Table~\ref{tab:cell_stats} ($D_{KAM}$). The results of the autocorrelation analysis
    are added ($D_{AC}$) - calculated as the geometrical averages of the $\phi$ and $\chi$ components from Table~\ref{tab:autocorrelation}. For average misorientation the literature values ($\Theta_C$) are compared to the cell analysis values ($\Theta_{KAM}$). }
    \label{fig:compare_size}
\end{figure}

In Fig ~\ref{fig:compare_size} we compare the average cells sizes and misorientation angles with the aforementioned TEM data. Given the different definitions of size, the TEM data should be multiplied by a stereological factor of $\pi \sqrt{pi}/4 = 1.4$ for a direct comparison.  In relation to the misorientation angle, on average the TEM values should be $\sqrt{3/2} = 1.22$ times larger, given the fact that TEM measures all three components of the orientation while the DFXM data relates to only two. Given also the difference between single crystal and polycrystal samples, the correspondence with the TEM data is seen as excellent. 

\subsection{Complementary study on a (111) single crystal with the TA within the inspection plane}

To confirm the finding from TEM that Geometrically Necessary Boundaries exist in this system, but only visible when using a different inspection plane, cf. Section~\ref{sec:compare_literature}, we here report on a supplementary DFXM experiment on a different specimen, but with the same sample material and dimensions as in the main text. Moreover, the x-ray set-up was essentially identical, but performed after the microscope was relocated to the new dedicated DFXM beamline, ID03 at ESRF. Deformed to an applied load of $\epsilon = 0.053$, the tensile axis was still [111], but the DFXM mapping was conducted using the (2-20) perpendicular to the tensile axis as diffraction vector. 

\begin{figure}[hbt!]
    \centering
    \includegraphics[width=0.75\linewidth]{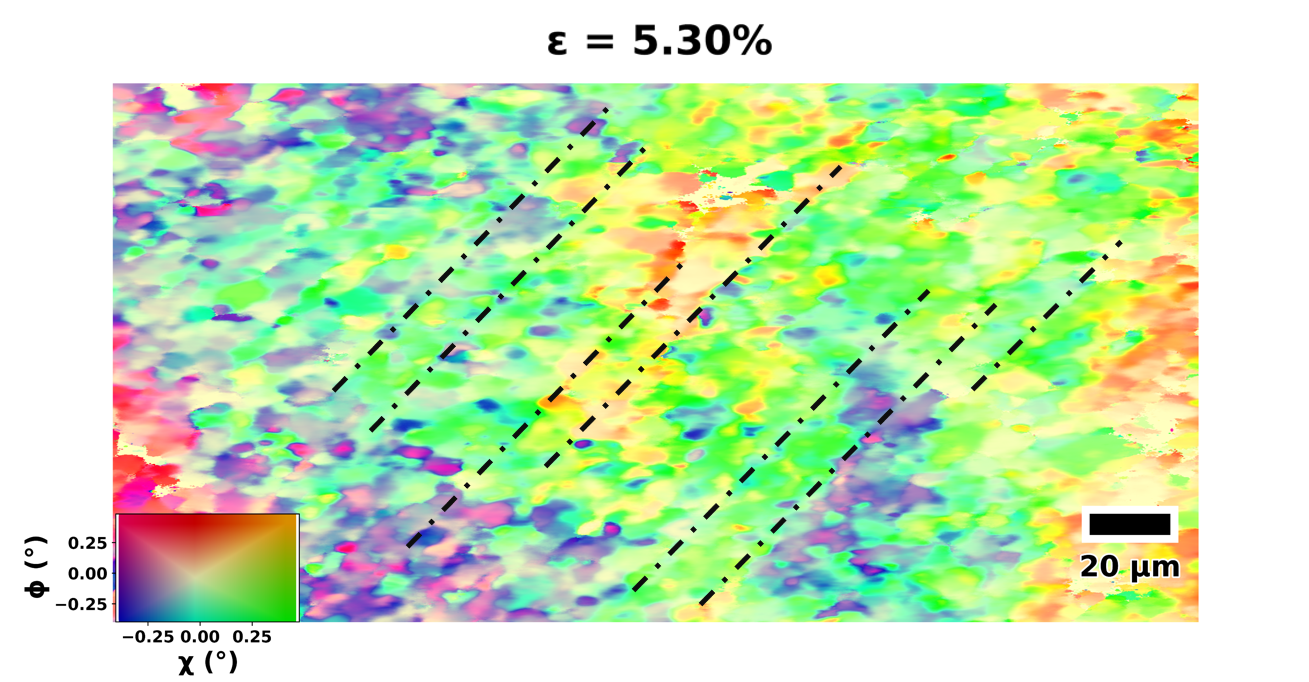}
    \caption{Orientation map for one inspection layer that is approximately orthogonal to the one in the main story for $\epsilon = 0.053$. Overlaid, in black, are guides to the eye, identifying GNBs.}
    \label{fig:OIM_other_plane}
\end{figure}

The resulting orientation map is shown in Fig.~\ref{fig:OIM_other_plane}. The orientation spread is similar but in this plane there is clear evidence of the GNBs.  

\vspace{2cm}
\subsection{Mathematics in relation to log-normal and chi distributions }
\label{sub:lognormal_math}
\hspace{1cm}

In this section we provide the parameterisations used in this work for the two distributions and we present the conditions for scaling.

\underline{The log-normal function},

$f(x)$ is a normalised function, parameterised as follows:
\begin{eqnarray}
    f(x) & = \frac{1}{\sqrt{2 \pi} }\frac{1}{\sigma x}  \exp \Big( -\frac{(\ln x - \mu)^2}{2\sigma^2} \Big) \label{eq:log-normal1}
\end{eqnarray}
With this parameterisation key properties of the distribution is given in Table.~\ref{tab:math_lognormal}. 

For $x \to \infty$ we have $f(x) \sim \frac{1}{x} \exp \Big( - (\ln x)^2 \Big) $.

\begin{table}[hbt!] 
\centering
\renewcommand{\arraystretch}{1.3}
\begin{tabular}{|c| c| c| }
    \hline 
     Mean & Median & Variance \\ 
     \hline
     $\exp (\mu + \sigma^2/2)$ & $\exp (\mu)$ & $\Big[ \exp (\sigma^2) -1\Big] \exp (2\mu +\sigma^2)$\\
    \hline 
\end{tabular}
\caption{Properties of log-normal distribution, with the parameterization expressed in Eq.~\ref{eq:log-normal1}. }
\label{tab:math_lognormal}
\end{table}

To create a log-normal distribution with a given mean $\mu_X$ and variance $\sigma_X$we 
generate cells using the random function (where $Z$ is a normal distribution)
\begin{eqnarray} 
X & = e^{\mu +\sigma Z} \; \textrm{with} \; \mu  = \ln \huge( \frac{\mu_X^2}{\sqrt{\mu_X^2 + \sigma_X^2}} \huge) \; \mathrm{and} \;  
   \sigma^2 = \ln \huge( 1 + \frac{\sigma_X^2}{\mu_X^2} \huge). \label{eq:find_mu_sigma}
\end{eqnarray}

Next, we establish the conditions for \textbf{scaling}. Consider two log-normal functions $f(x,\mu', \sigma')$ and $g(x,\mu, \sigma)$. Then these exhibit scaling with a factor
$k$ if and only if
\begin{align}
f(x,\mu', \sigma') & = k g(kx,\mu, \sigma).
\end{align}
For this to be true at all x we have 
\begin{align}
\mu' & = \mu - \ln k; \hspace{1cm} \sigma' = \sigma
\end{align}

\hspace{1cm}

\underline{The chi function}

$f(x)$ is a normalised function:
\begin{eqnarray}
    f(x;\sigma,k) & = \frac{1}{2^{k/2-1} \Gamma(k/2)}  \big( \frac{x}{\sigma} \big)^{k-1}  \exp(-\frac{1}{2}(\frac{x}{\sigma})^2 )  \label{eq:chi1}
\end{eqnarray}
Here $k$ represents the number of degrees of freedom.  For $k=2$ it becomes the Rayliegh distribution, for $k=3$ the Maxwell-Boltzmann speed distribution.  With this parameterisation key properties of the distribution is given in Table.~\ref{tab:math_chi}.

\begin{table}[hbt!] 
\centering
\renewcommand{\arraystretch}{1.3}
\begin{tabular}{|c| c| c| c|}
    \hline 
     & Mean, $\mu$ & Median & Variance \\ 
     \hline
     General $k>0$ & $\sqrt{2} \frac{\Gamma((k+1)/2)}{\Gamma(k/2)} \sigma$ & $ $ & $k - \mu^2$\\
     Rayleigh: $k=2$ & $\sqrt{\pi/2} \; \sigma$ &  $\sqrt{2 \ln 2}\; \sigma$  & $(4-\pi)/2 \;\sigma^2$\\ 
    \hline 
\end{tabular}
\caption{Properties of chi distribution, with the parameterization expressed in Eq.~\ref{eq:chi1}. }
\label{tab:math_chi}
\end{table}

To create a chi distribution with a given mean $\mu_X$ and degrees of freedom $k$ we
generate cells using the random function (where $Z_i$ are  normal distributions in the $i=1\ldots k$ directions)
\begin{eqnarray} 
X & = \sqrt{\sum_{i=1}^k Z_i^2}  . \label{eq:generate_chi}
\end{eqnarray}

All chi distributions with same k scale automatically.

\end{document}